# Construction of 3D-Ferroelectric Polarization Microstructure and Detection of Polarization Invariants under Induced Flexoelectric Strains

*Sabarigresan Murugan*[1], *Vaishnavi S M*[1], *Ranjith Ramadurai*[1]*

[1]Department of Materials Science and Metallurgical Engineering, Indian Institute of Technology Hyderabad, Kandi, Sangareddy, Hyderabad. 502285.
**Email:** *ranjith@msme.iith.ac.in*
ORCID ID: https://orcid.org/0000-0003-2991-0027

**Funding:** DST-SERB India for the funding (Grant No.: DST/SERB/EMR/2017/003159/MMM), SATHI-CISCoM center at IIT Hyderabad funded by DST India, SR/SATHI/2022/247.

**Abstract**

Ferroelectric thin films that are heterophased and polycrystalline possess strain-sensitive piezoelectric behavior. However, a direct insight into polarization orientations within a given grain and how mechanical stresses reconfigures the ferro elastically coupled polarization orientations remains ambiguous. A polarization component-resolved imaging in a Piezoresponse Force Microscope (PFM) was developed in combination with correlative structure and phonon studies that provide insights into grain orientation, domains and bending stress-driven phase transitions. The method correlates the polarization invariant and the respective grain orientations present underneath. The technique facilitates an experimental inverse model approach to determine crystallographic grain orientation from the ferroelectric domain. A three-point bending stage introduces a flexoelectric strain in $Ba_{0.85}Ca_{0.15}Zr_{0.1}Ti_{0.9}O_3$ (BCZT)thin films and simultaneous polarization imaging. This novel technique potentially captures the formation of new polarization invariant of a monoclinic phase that appears under high pressures. The experimental findings pave way for development of inverse modelling of heterophased and polycrystalline systems.

## 1. Introduction

Ferroelectric materials are indispensable to micro/nano-electromechanical systems (MEMS/NEMS).[1] They are known to possess a switchable spontaneous polarization with a simultaneous ferroelastic strain that offers superior electromechanical coupling.[2] The inter-relation between the polarization orientation across domains with varying configurations gives rise to exotic phenomena, like vortices, polar skyrmions, and novel hybrid domain patterns.[,3,4] The interplay between the electric dipole ordering and their alignment with the respective adjacent dipoles and domains opens up interesting fundamental phenomena like novel antiferroelectric ordering and polar meron structures under specific conditions.[5,6] The superior electromechanical response of the morphotropic compositional ferroelectric materials (eg. $Ba_{0.85}Ca_{0.15}Zr_{0.1}Ti_{0.9}O_3$ (BCZT)) originates from comprising both intrinsic and extrinsic factors.[7,8] However, in practical applications, the polycrystalline ferroelectric thin films are predominantly used in spite of the exotic phenomena of polar ordering observed in epitaxial thin films.[5,6] The compositional heterogeneity offered by morphotropic compositions and the associated heterophased character adds up to the complications of studying the polarization ordering in these systems. The experimental observation of the polarization orientations in a polydomain structure, especially with a heterophased character, remains completely understudied.[9,10,11,12,13,]. In this study, we attempt to develop a method of imaging the various polarization variants present in heterophased BCZT thin films, which leads to a correlation between the domain orientations present on the top of a polycrystalline microstructure. The technique is also verified to provide grain orientation information from the observed polarization images, which is generally performed using diffraction techniques like EBSD.[14]

Ferroelectric solid solutions of BCZT with a morphotropic phase composition was chosen in this study due to their delicate energy landscape that facilitates the polarisation rotation and ferroelastic switching[15,16,17,18,19,20]. It also possesses relatively larger piezoelectric coefficients ($d_{33} \geq \sim 620$ pC/N) comparable to those of conventional lead-based piezoelectrics[21,22,23]. The morphotropic composition of polycrystalline BCZT thin films naturally leads to heterophase conditions and offers an excellent case study for experimentally observing a correlation between the existing domain pattern and the underlying microstructure.[24,25,26]

Through 3D PFM (piezoresponse force microscope) image reconstruction and structural analysis, we correlate the existing coupling between crystallographic orientation, domain topology, and electromechanical responses, which are key parameters in determining the efficiency and reliability of MEMS devices. In this work, we attempt to establish a framework for designing and providing plausible experimental inputs for inverse modeling of ferroelectric thin films with tunable microstructure and domain patterns, leading to aspired device performances for next-generation sensors and actuators. The study also includes the development of a 3-point bending stage that enables *in-situ* application of uniaxial bending stress and study correlative properties under piezoresponse force microscopy, X-ray diffraction, and Raman spectroscopy with different flexoelectric strain conditions.

## 2. 3D Polarization Reconstruction: Decoding Phase Heterogeneity via Orthogonal PFM

The widely utilized “Vector PFM” method has been a potential tool and facilitates resolving the polarisation components for epitaxial thin films with known crystallographic orientation.[27,28] The Fig. S1 in the supplementary shows the PFM image of BCZT thin films, which clearly shows that the domains are across various grain boundaries (highlighted on S1). The neighbouring grain orientation is crucial for the crossover of the domain across grain boundaries such that the energy of the domain wall formation surpasses the energy required to overcome the grain boundary barrier. Especially in ferroelectrics with MPB, where multiple phases coexist, strain transfer across grains, depending on the neighbouring grain orientations, can significantly influence the domain and domain wall dynamics. EBSD mapping in the same location resulted in blurred Kikuchi lines due to overlapping signals from neighbouring grains (S2). However, there are limitations in extending this commercial package to polycrystalline thin films. In this work, we develop a methodology to establish an angular relation between the polarization components, the polar direction of the ferroelectric phase, and the crystallographic orientations of the grains that comprise both tetragonal and orthorhombic phases of BCZT. Similar studies of 3D reconstruction have been reported for epitaxial films in single-phase materials[29,30]. To overcome the limitations and/or complications involved in heterphased polycrystalline thin films, we used a PFM operating under high vacuum conditions with the hardware that enables us to trace back to the same location of a given sample, effectively minimizing the error via pixel-by-pixel correlation across the dataset. The two-step process used in this work records the vertical (V) and lateral (L0) signals of PFM, followed by an additional lateral imaging (L90) by rotating the sample by 90°. Later, the signal is coded with the colors representing the polarization orientation of a given location with respect to the cantilever. (see color scale bar (Figure 1) with angular information). Further implementation was done with a quantitative analysis workflow in MATLAB to visualize the "real" piezoresponse (R), which is detailed in the methods section. The MATLAB script then processes the data on a pixel-by-pixel basis, and for each point of data, it generates a 3-bit binary signature [V, L0, L90 – Table 1 from the methods section]. The output of the program is a map that identifies each location to a specific ferroelectric polarization orientation within a single or multiple grains.

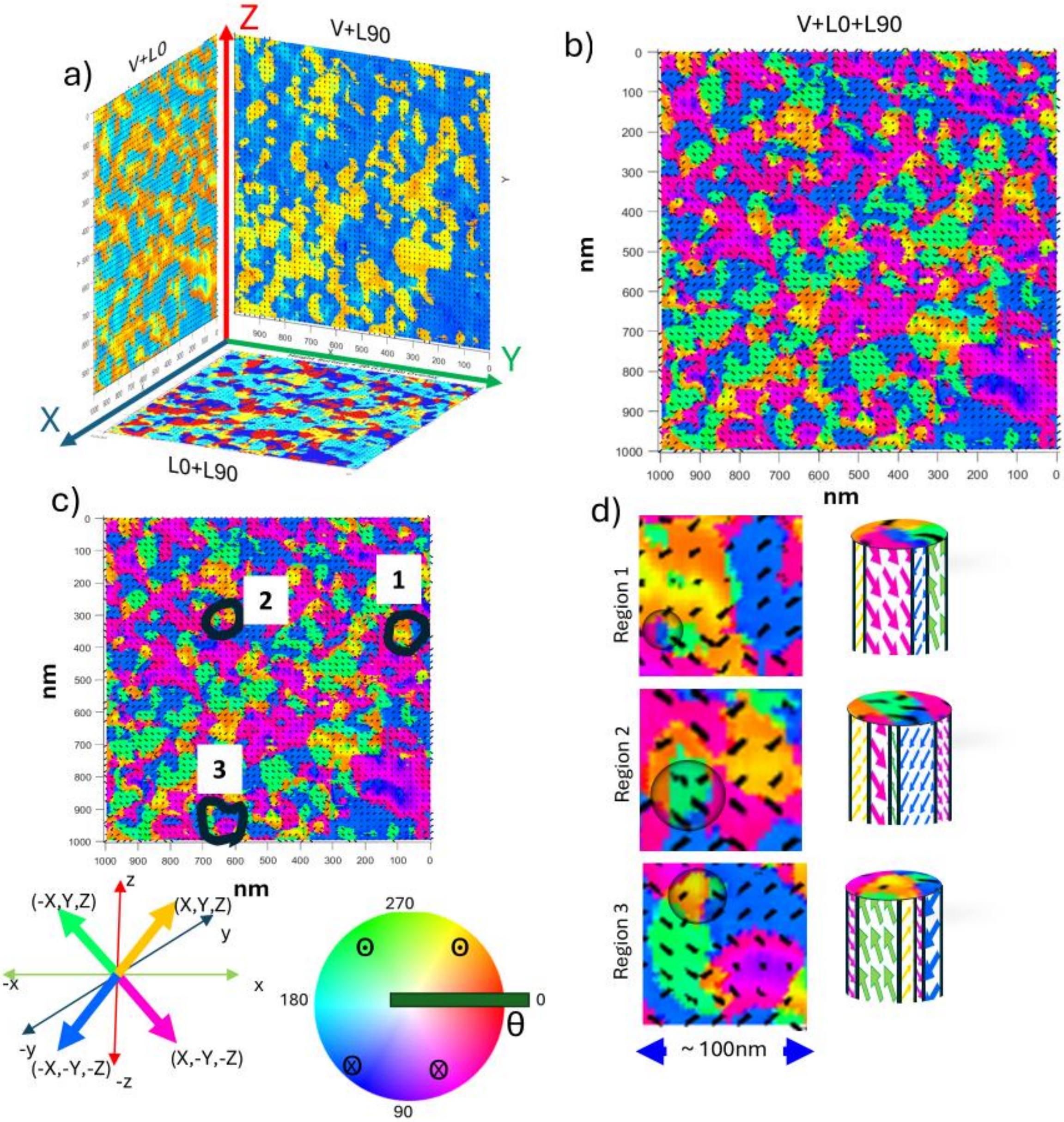

***Figure 1.*** *Three-dimensional polarization mapping and domain topology of BCZT thin films.* ***a)*** *the reconstructed 2D polarization vector maps across the* ***XZ, YZ, and XY*** *planes and* ***(b)*** *and 3D resultant vector map* ***(c)*** *for a polycrystalline BCZT film, showing a transition to a more disordered, fine-grained domain structure.* ***c,*** *3D resultant map of the polycrystalline film with three representative regions (1, 2, and 3) highlighted to denote areas of high topological complexity.d, High-magnification views of the regions identified in* ***(d)****Colored arrows indicate*

*the local orientation of the polarization vectors, revealing the presence of domain walls configuration within a 100 nm scale.*

Figure 1(a) shows the piezoelectric response maps across the XZ (vertical and lateral response (L0) of the initial sample mount), YZ (vertical and lateral response (L90) of the sample rotated by 90°), and XY planes. The images were obtained from the resultant phase angles from both the vertical and lateral responses. More details of the construction of the figures could be found in the methodology, and Figure 1(b) shows the resultant components obtained from all planar projections of 1(a). It represents the 3D projection of the effective polarization orientation on the grain morphology of the heterophased polycrystalline film. The method developed in this manuscript facilitates a pixel-by-pixel matching of the reconstructed polarization orientations with the morphology of the sample. The details of the minute resultant polarization components over a morphological region could be found in Figure 1(c). The regions (1, 2, and 3) highlighted in Figure 1(c) show the domain complexity in a polycrystalline BCZT film. Figure 1(d) shows the magnified view of the highlighted regions of 1(c). The resultant polarization orientation underneath the color map is represented beneath the region. It is evident that certain polarization components are emerging out-of-plane (⊙ in the color wheel) and certain components are pointing below the plane (⊗ in the color wheel). However, the equivalent vector arrows calculated and drawn beneath the color map evidently show head-to-tail (green and blue or yellow and pink) configurations present adjacent to each other (region 1). It is also evident that two regions of head-to-tail configuration are bridged with a 90° domain region, leading to a head-to-head configuration (pink region). (e.g., pink arrows as seen in region 2 and region 3). Thus, the method developed enables us to visualize extremely tiny regions of polarization orientations that are otherwise buried under the average lateral and vertical responses of the PFM studies. Most importantly, it also suggests that recent observations like hybrid antiferroelectric-ferroelectric domain walls could be present in heterophased, polycrystalline systems to accommodate the competing strain and/or depolarization energies between the coexisting polar phases. Such details would otherwise not be captured under dual contrast images, most importantly for heterophase materials, as shown in supplementary figure S3 obtained from the same region. Such an analysis of domain configuration could lead to analyzing the polar order parameters like polar skyrmions, vortices, and their respective polarization orientation [31]. This method can propel the analysis from being a standard ferroelectric materials technique by enabling 3D reconstruction in high resolution, only limited by the probe tip diameter, where changes from one transitional antiphase state to another can be visualized. To eliminate the ambiguity on the influence of morphology, shadowing the piezo forces experienced by the cantilever, the 3D surface overlay of the grain and the 3D resultant with different polar vectors are provided in the supplementary figure (S3).

## 3. Importance of domain configurations on nanoscale polarization switching

The piezoelectric coefficient($\mathbf{d_{ijk}}$ ) a third-rank tensor, exhibits strong dependence on the crystal structure, domain structure, spontaneous polarization ($\boldsymbol{P_s}$) and the strain present in the material. The coupling between all of them determines the performance of the material. The energy landscape in BCZT for varying polarization orientations can vary locally where there are multiple domain configurations. [14]

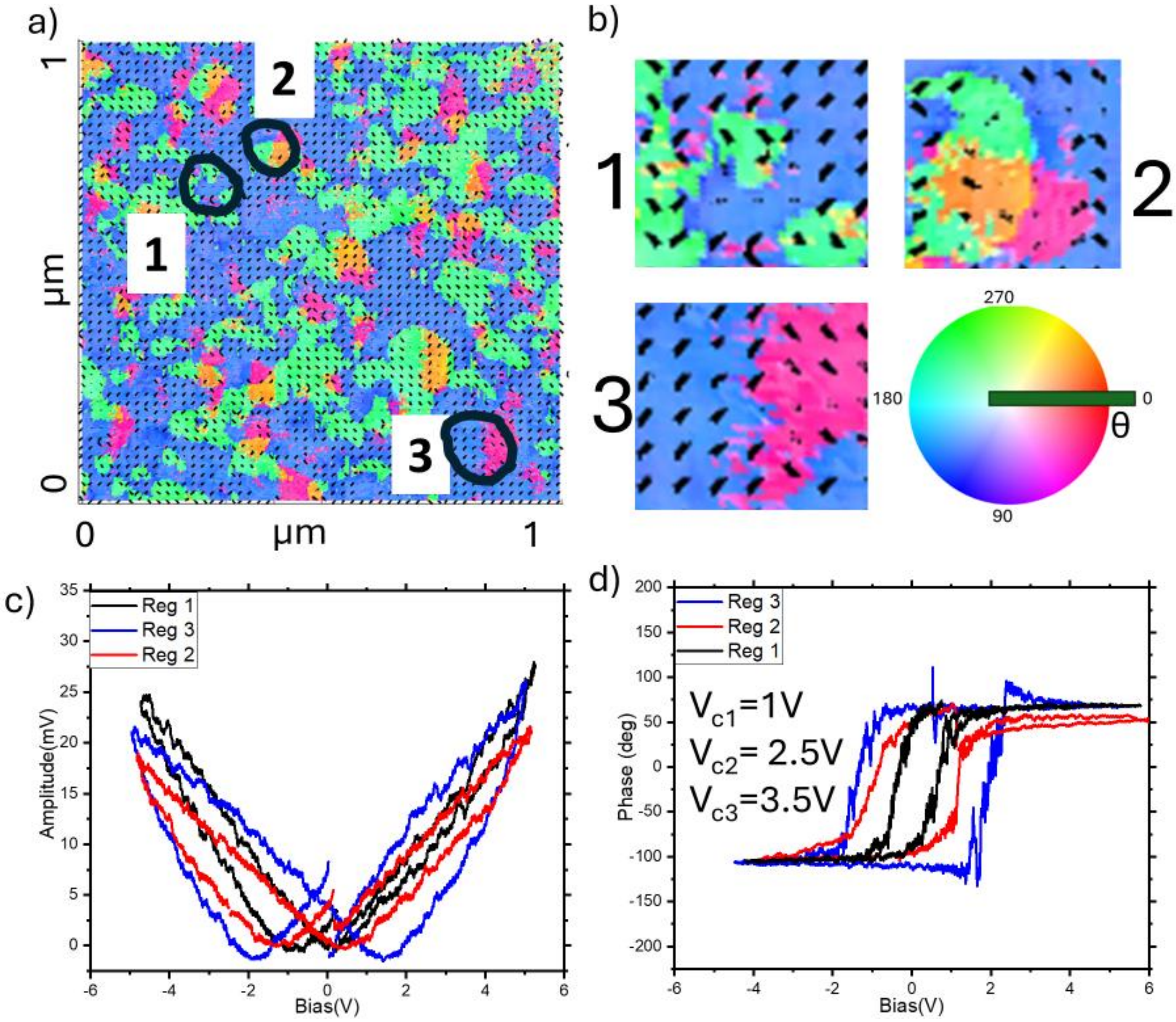

***Figure 2. a, b.*** *3D polarization map of BCZT film and zoomed regions;* ***c****. PFM amplitude vs. bias and* ***d****. phase vs. bias of different regions.*

Figures 2(a) and 2(b) highlight three distinct domain configurations in the heterophase polydomain thin film. The switching spectroscopy studies evidently reveal that the device parameters, like coercive voltage, area under the loop, etc., are sensitive to the underlying polarization microstructure. Thus, highlighting the importance of the polarization microstructure for developing physics-informed models like digital twin developments for ferroelectric-based devices. The presence of different polarization microstructures could also be evidently observed by the intermediate phase states observed in the time-domain switching studies (S5), confirming that local domain configurations dictate the switching pathway. Similar intermediate PFM phase signals have been observed during polarization reversal in PMN-PT, where they are described as a 'dense domain structure' arising from localized elastic clamping. [30]. These states represent a transition to a lower-symmetry monoclinic phase, which acts as a structural bridge during the strain-induced transformation [32].

4. **Identification of Grain Orientation using Piezoresponse Force Microscope.**

The study naturally led to establishing a relation between the polarization components and their angular relations with the cantilever, along with the underlying grain orientation of a known crystal structure. For example, consider a [001] grown epitaxial ferroelectric thin film with a tetragonal structure, whose spontaneous polarization is parallel to the growth direction. This leads to a higher vertical piezo response with zero lateral responses depending upon the mosaicity of the film.

Correlation between the specific polarization axis dictated by crystal symmetry and the piezo response behavior facilitates to develop a reverse approach to narrow down the possible grain orientation of a polycrystalline thin film. The only alternative to make such a meaningful correlation between grain orientation and the respective crystal structure is a very few microscopy techniques, like Electron backscattered diffraction (EBSD). However, it possesses certain limitations (Supplementary S2) to utilize it for fine-grained (~ 20-60 nm) thin films due to the overlapping of Kikuchi lines, weak signals resulting in improper indexing of crystallographic orientations[33,34]. If similar studies are employed under transmission electron microscopy, the samples are forced to experience ion bombardments, which could alter the delicate strain energy balances possessed by the BCZT type of thin films whose $T_C$ is close to room temperature.

In this study, by employing the orthogonal reconstruction as explained in Figure 1, a novel approach was developed to reverse engineer the property information into structural information and predict the orientation of grains.

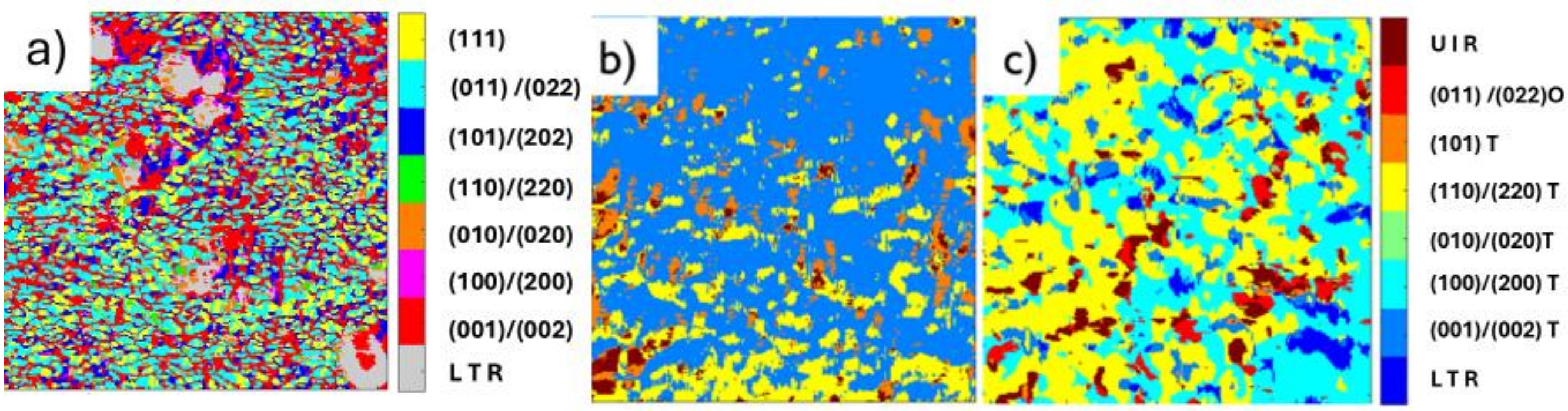

***Figure 3*** *a) shows the orientation map re-created using 3D PFM of polycrystalline BTO, and b) shows the orientation map of the epilayer of BCZT film deposited on a* $DyScO_3(001)_{pc}$ *substrate. c) shows a polycrystalline map.*

Figure 3(a) shows grain mapping of a $BaTiO_3$ polycrystalline film (used as a standard) with a tetragonal phase; 3(b) shows the resultant orientation map of epitaxial BCZT with a known mixed phase of tetragonal and orthorhombic; [35] 3(c) shows the polycrystalline BCZT orientation map. However, there is a fundamental ambiguity concerning the (001) tetragonal and (111) rhombohedral orientations since both grains produce only a strong vertical piezoresponse, making them identical to our 90° orthogonal reconstruction. This isn't a problem that can be solved by rotating to a different angle; a (111) rhombohedral grain simply has no lateral PFM component to measure. Furthermore, a distinct challenge arises in differentiating between (111) and (110) tetragonal grains, as their piezoresponses are nearly identical. Due to this lack of signal contrast, these orientations cannot be reliably distinguished based on magnitude alone. To further confirm, the same was carried out in $BaTiO_3$ polycrystalline films (methods Figure 6a), where we can use the crystallographic information to predict the direction of low-angle grains that are closer to (100) & (110) planes. Whereas if we know more about the sample's crystallographic information, like the epitaxy-oriented films and single-phase polycrystalline films, we can reduce the unknown variable and increase the accuracy of indexing. For an analyzed area of 1 $um^2$, we have identified the area distribution percentage of BTO polycrystalline thin film as <001> is 36.4%, <110> is 40.2%, <111> is 19.1%, and LTR is 4.3%. For BTO-like single-phase materials, this becomes very subtle, as we know the crystallographic planes, which correlate with XRD. However, the complexity increases significantly for heterophase polycrystalline thin films characterized by local energy variations. In such systems, determining phase distributions via XRD is often impossible without intensive Rietveld refinement. Furthermore, XRD lacks the spatial resolution required to locally probe a specific region of interest. Our method effectively addresses these limitations, providing high-resolution orientation mapping.

With the prior knowledge about the grains that are present in the X-ray diffraction studies, the angular relation between the polar axis and the grain orientation axis is calculated, including the possibility of a heterophase, and later, the crystalline grain orientation underneath the polarization orientation could be predicted. In order to verify the performance of the developed method, experiments were performed on a sample with known crystallographic orientations, such as (001) and (110), which is provided in the supplementary information (S4).

## 5. Flexoelectric strain-induced polarization switching and detection of new polarization invariants

To understand the potential of the construction of the 3D polarization microstructure, the studies were carried out on samples that undergo extrinsic stress conditions similar to the prestrained conditions of MEMS devices. This facilitates visualizing the changes experienced by the polarization microstructure under certain device operating conditions. To mimic the prestrained condition, a novel 3-point bending stage was designed and fabricated as shown in Figure 7(a) of the methods section. It was designed to fit across multiple equipment for correlative *in-situ* and *operando* studies under flexural stress, which can offer a maximum load of 98 N (with contact pressure approx. 2.3 GPa; refer to the methods section). Other methods of flexoelectric or uniaxial bending stress-induced effects on studying ferroelectric domains and electromechanical response involve using a mandrel-induced radius of curvature or the preparation of free-standing ferroelectric membranes supported by flexible substrates[36,37] whereas the *in situ* bending stage offers an advantage over these methods since it allows direct mounting of the polycrystalline BCZT thin film-substrate assembly to perform characterization with different probing techniques.

The PFM images (supplementary figure S8) as obtained show a quantitative insight into the domain switching under a progressive bending load. The lateral piezo response peaks at 84N due to ferroelastic reordering, but at 98 N, the strain-induced structural disorder declines the peak in-plane response. Conversely, out-of-plane increases through the bending but saturate after 84 N. However, the pristine and final domain state proves that there is a significant irreversible change due to flexoelectric strain. This provides direct, nanoscale proof that the stress-induced polarization is non-volatile, leaving a substantial remanent "mechanically poled" state. In these MPB-based thin films, this permanent shift likely reflects a phase transition facilitated by a flattened free energy, where the mechanical load drives the material into a new, stable polarization state that persists even after the stress is removed. The PFM images shown in S9 (a) show how the domain reorients with initial loading conditions, which might be a manifestation of new polar states, possibly orthorhombic or rhombohedral phases. To transform the as-acquired PFM image into interpretable resultant orientation maps, our method of reconstruction is utilized.

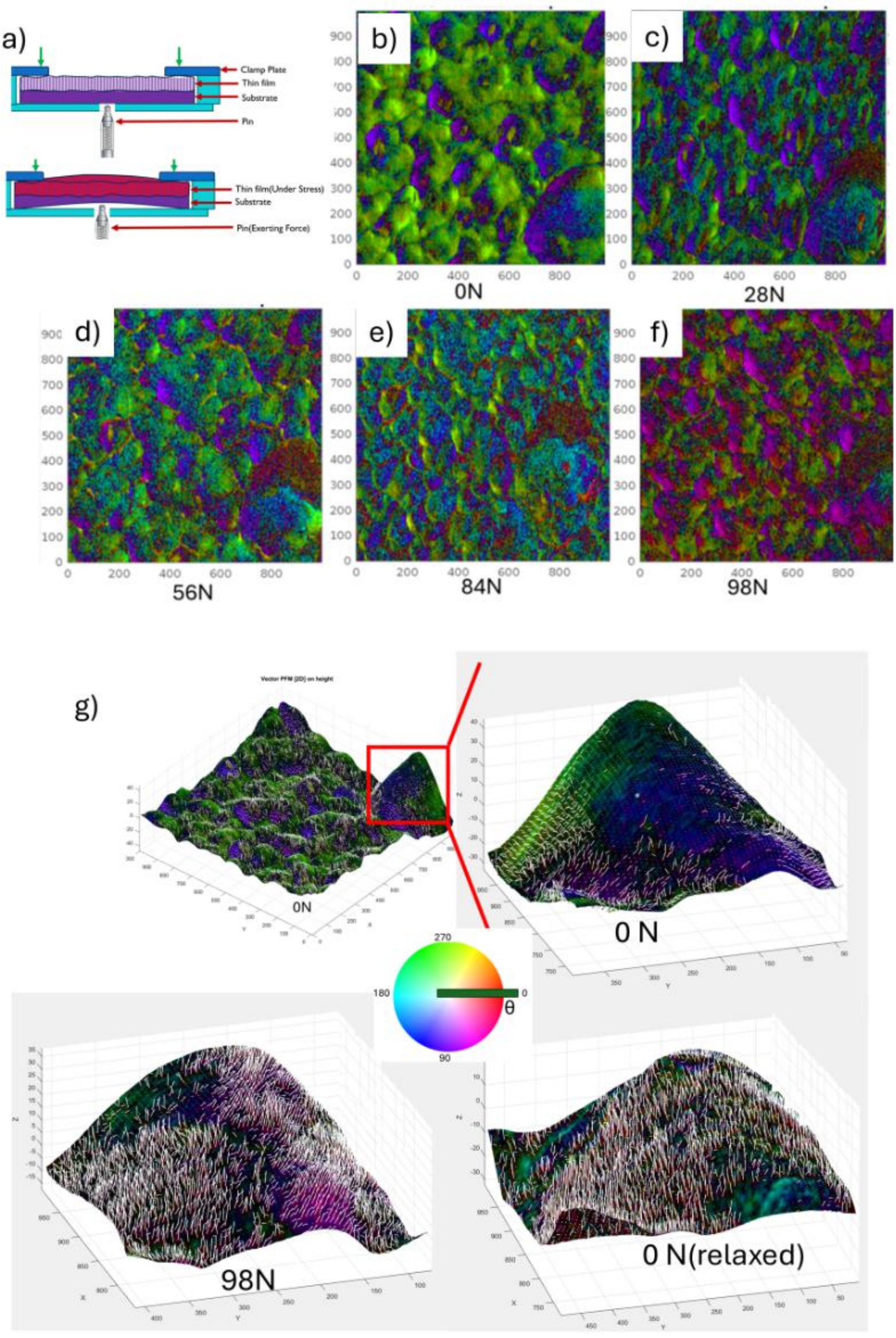

***Figure 4.*** *a shows the schematic of 3-point bending; b, c, d, e, and f show the reconstructed polar map of the BCZT film under bending loads; g and h show the 3D overlaid polar map with vector arrows representing polarization rotation (with deliberately reduced density of arrows to facilitate visualization for readers).*

The obtained PFM amplitude indicates an overall enhancement of piezoresponse with bending, but it is insufficient to resolve domain wall relationships. In contrast, Figure 4 (a) to (e) shows resultant 2D PFM maps reveal that the pristine film is dominated by two contrasts separated by ~180°, consistent with a well-defined tetragonal phase with simple up/down domains. Under bending stress, this configuration destabilizes, and intermediate phase contrasts emerge, signifying polarization changes and the activation of non-180° domain orientations. 2D reconstruction from combined vertical and lateral PFM signals directly visualizes the switching pathway, revealing progressive polarization change and the formation of tilted domain states. The polarization changes indicate irreversible changes, demonstrating changes in domain orientation driven by mechanical bending. The PFM amplitude loops (S10) exhibit a characteristic butterfly loop, with a notable reduction in peak electromechanical response under a 98N bending. This suppression, coupled with the incomplete recovery observed in the state after a full bending cycle, suggests that external stress induces a significant domain clamping and a shift in the local coercive field, characteristic of stress-mediated ferroelectric tuning. The observed reduction in coercive voltage under mechanical load suggests a lower energy barrier for domain reversal, potentially beneficial for low-power device operations.

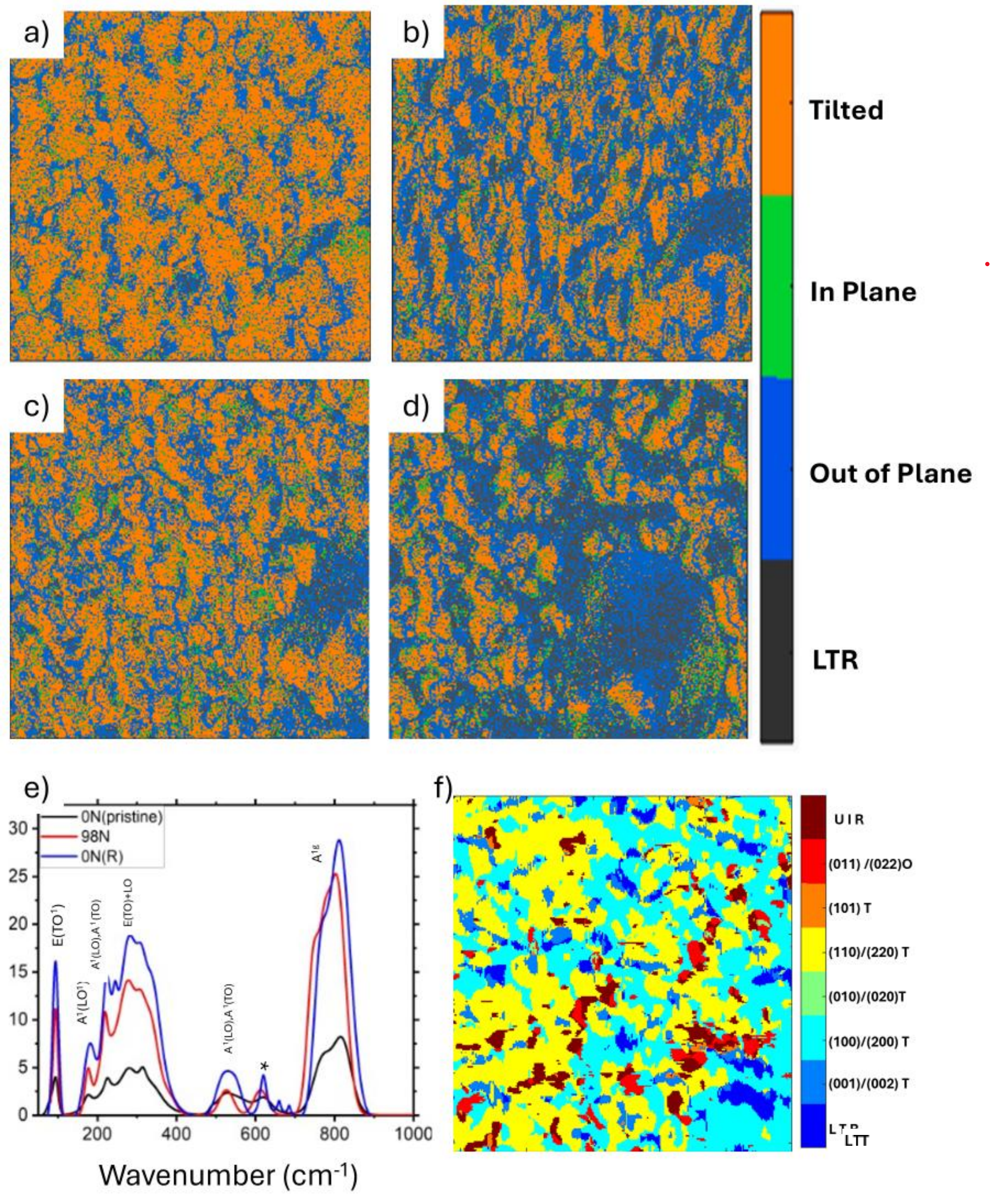

***Figure 5 a, b, c, and d are*** *2D orientation maps under bending conditions 0 N, 28 N, 98 N and 0 N(return),* ***e)*** *Raman spectrograph with in-situ bending, and* ***d)*** *shows polycrystalline film orientation map after bending.*

In the pristine state, the PFM response is seen in Figure 5(a) with a more tilted response where both in-plane and out-of-plane responses are present, indicating more of (110) and (100); as we go through the loading, we can see that the orientation of the response switches to more out-of-plane responses, making the polarization switch from their parent phase. Correlating that to structural studies through Raman spectra, Figure 5 (e) also provides strong evidence of an irreversible, stress-induced phase transformation in BCZT. Initially, the 0 N spectrum exhibits a tetragonal-rich state; however, under 98 N bending, rhombohedral signatures are suppressed while tetragonal/orthorhombic modes intensify. The emergence of a new mode near 630 $cm^{-1}$ suggests a strain-stabilized low-symmetry bridging phase, and the sharpened remanent spectrum (0 N(R)) confirms permanent ferroelastic mechanical poling into a new stable phase. The new mode emerging at ~ 630 $cm^{-1}$ suggests a strain-stabilized low-symmetry bridging phase analogous to high-pressure monoclinic states reported in literature. [38,39] The remanent 0 N(R) spectrum further confirms irreversible ferroelastic mechanical poling, which corroborates the PFM observations. Figs. 3(b) and (c) exhibit regions with a distorted response that cannot be definitively classified as tetragonal or orthorhombic, presented as UIR, suggesting an intermediate distortion between them. This signature, typically found at the predicted phase transitions of the parent BTO compound, hence provides new evidence of a possible monoclinic distortion at room temperature without the need for external thermal or mechanical stimuli. The FWHM and wavenumber evolution (S11) with bending load reveals the mechanism of stress-induced transformation, whereas O/T modes remain permanently broadened, indicating irreversible ferro elastic locking in the O-phase, where low energy levels facilitate strain-induced transitions at ambient conditions even though they possess a similar $T_c$. Correlative GIXRD also confirms R-phase ordering via (111) splitting at 28 N and remanent O-phase modification through persistent (110) changes, while the R-phase distortion remains elastic and reversible (S13).

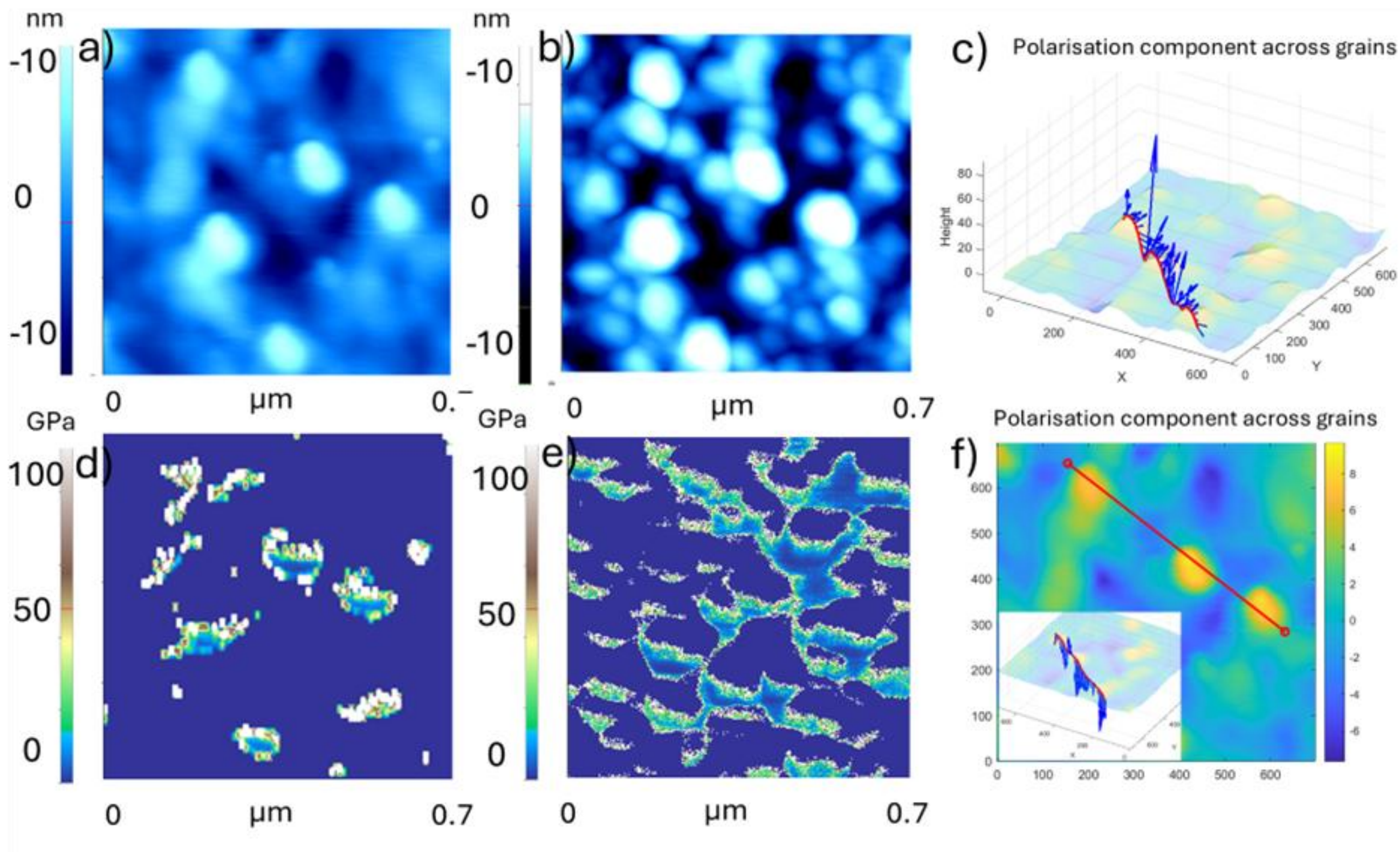

***Figure 6, a** and b, shows height; d and e show modulus maps; **c** and **f** show line profiles of resultant piezo response before and after heat treatments.*

Nano-mechanical studies were done on the same sample after the bending cycle ( Figure 6(a and d)). Then the film was thermally cycled to 110 °C, held for 1 h, and cooled at a slow cooling rate of 3°C $min^{-1}$ to avoid thermal mismatch cracking, after which the same region was re-examined for nano-mechanical investigations ( Figure 6(b and e)) to shed more evidence on the ferroelastic domain presence and to reverse the induced phase transition by bending, where we observe a significant change in modulus and domain distribution indicating the emergence of ferroelastic domain variants when bending as a strain relaxation mechanism. Residual stress analysis S12 (a and b) also confirms that as-grown films are clamped in a tetragonal state by substrate-induced strain from thermal and lattice mismatch with Pt/Si, suppressing the MPB behavior. The flexural stress offered by the bending stage facilitates a strain relaxation leading to the formation of orthorhombic ferroelastic variants from the existing tetragonal state via monoclinic distortion, indicating a transition towards bulk-like MPB nature.

## 6. Conclusions

In this work, we establish a direct nanoscale correlation between grain orientation, domain configuration, and flexural strain in lead-free BCZT thin films. The technique developed for orthogonal 3D polarization mapping facilitates visualization of the domain patterns overlaid on the morphology with their resultant polar components. This approach will aid in eliminating morphological ambiguity and reveal the transition to disordered, fine-grained domain structures that standard PFM analysis does not capture. These analyses will be a primary tool for decoding phase heterogeneity and identifying the real projection (arrows in the map) of the piezoresponse across different crystallographic planes. By inverse modeling the anisotropic property of piezoresponse into structural data, we can predict grain orientations in fine-grained films where traditional techniques like EBSD are limited. Finally, by integrating a custom three-point bending stage, we captured the dynamic evolution of these polarization states under induced flexoelectric strains. We observed a sequential transformation of BCZT thin film from a tetragonal-dominant state to an irreversible ferroelastic orthorhombic reordering. We have evidence to believe that the unidentified regions (UIR) are possible monoclinic distortions. This bending condition and the PFM orientation map are instrumental in identifying new polarization variants, such as the existence of room-temperature monoclinic distortion in BCZT, which acts as a structural bridge state during strain-induced transformations. Strain gradients can be utilized to pre-set domain orientations, offering a powerful method for optimizing the performance and power consumption of next-generation low-power devices.

## 7. Methods

### 7.1. Microstructural tunability under varying deposition conditions:

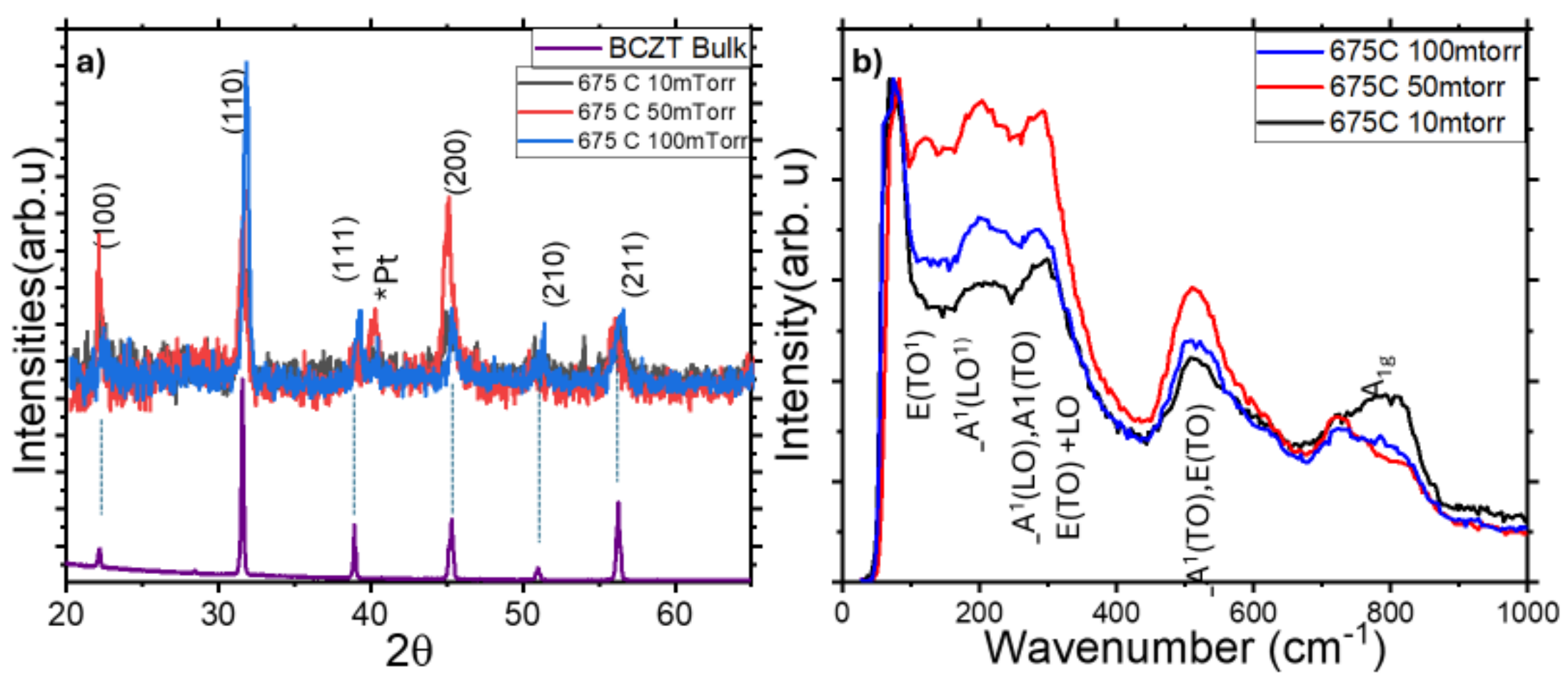

***Figure 7 (a)** XRD patterns of BCZT polycrystalline thin films deposited at 675 °C under varying partial pressures of oxygen. Variations in peak intensities indicate changes in preferred orientation and overall crystallinity with pressure. X-ray diffraction pattern of bulk BCZT pellets is used as a reference **(b)** Raman spectra of BCZT thin films grown at 675 °C at 10, 50, and 100 mTorr. The evolution of the vibrational modes reflects modifications in lattice order and structural distortion as a function of deposition pressure.*

BCZT ceramics with morphotropic composition were synthesized via the conventional solid-state reaction route and were sintered to dense pellets, which served as a target material for pulsed laser ablation and deposition (PLD). The target was placed in a high vacuum chamber, and the thin films were deposited using a KrF pulsed excimer laser, ($\lambda$~248nm) as the ablation source. Thin films of BCZT were fabricated on commercially available <111> platinum-coated silicon ($Pt/TiO_2/SiO_2/Si$ (100)) substrates (MTI Corp.). The BCZT thin films were grown at substrate temperatures ($T_S$) of 650, 675, and 700°C under an oxygen partial pressure of 10, 50, & 100 mTorr for each temperature with an optimized thickness of 200 nm. High Resolution X-ray Diffractometer (HRXRD) (Bruker AXS D8 Discover) was used to study the structural details of the thin films (Cu source ($\lambda$=1.541 Å)) in grazing incidence (GIXRD) mode. Piezoresponse force microscope (PFM) was used for imaging the ferroelectric domains, switching studies, and the respective 3D analysis using High vacuum PFM (HiVac and NX-10 for preliminary studies, Make: Park Systems). Raman spectra were studied with a 20-point average at each location using Witec confocal Raman equipment with a green laser (532 nm) for all the results presented.

The diffraction pattern of BCZT thin films and bulk ceramics shows that the thin films were phase pure and no impurity phase was detected. Though all the films exhibited phase purity, the BCZT films grown at a $T_S$ of 675°C and 50 mTorr were used for further studies due to their well-defined intensity ratios across various (*hkl*) planes. The observations were also corroborated by strong Raman signals (Figure 7 b).

### 7.2. Piezoresponse Force Microscope (PFM):

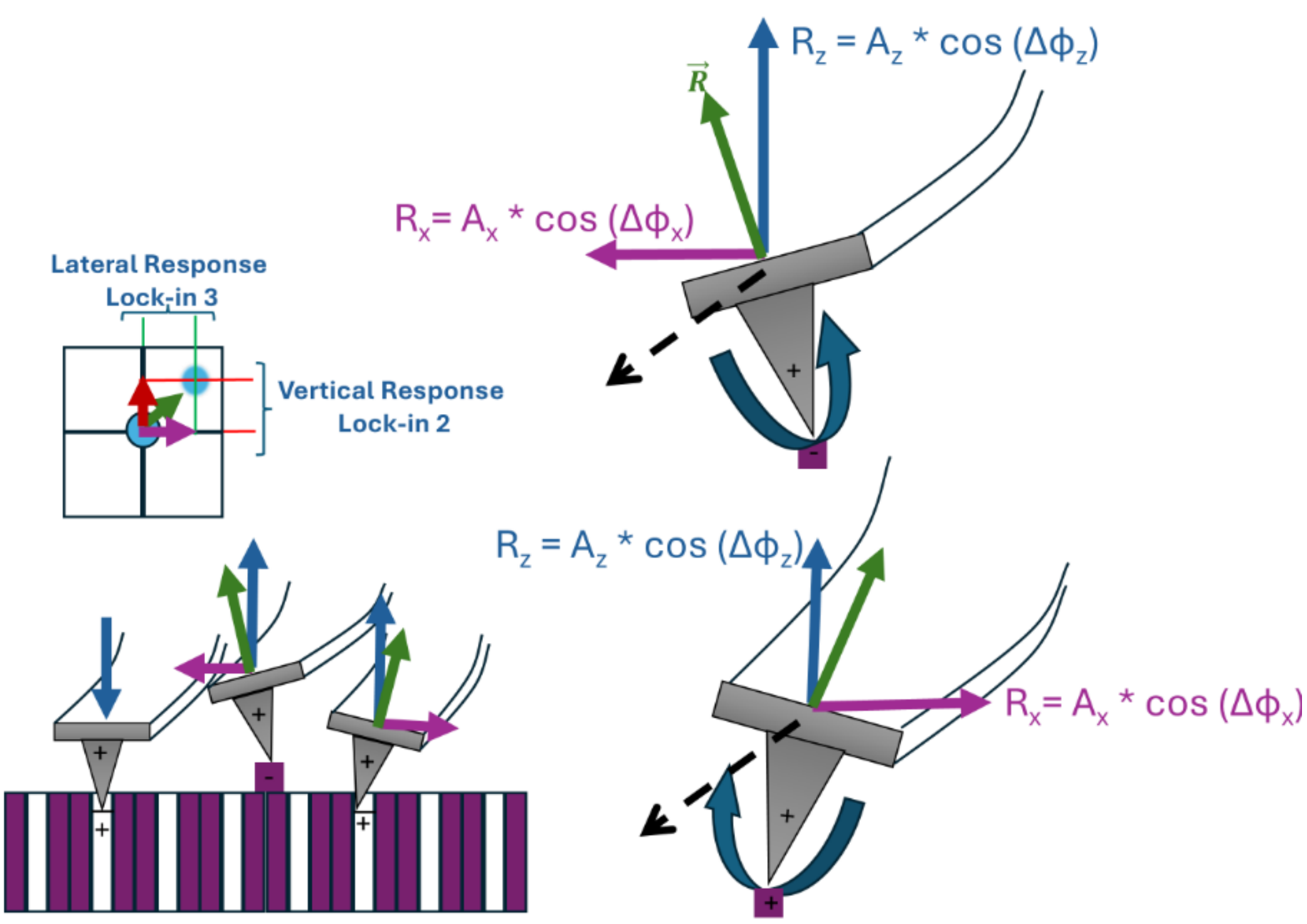

***Figure 8*** *shows a schematic of the operation principle of a PFM detailing the interconnection between the lock-in amplifier and Z servo, which work in tandem to capture i)topography, ii)vertical phase, and respective amplitude; iii) lateral phase and respective amplitude. The right image shows the interaction of the tip with a ferroelectric sample and the components of the response forces.*

The principle of PFM experimental architecture, illustrated in Figure 8, shows a conductive cantilever as a localized top electrode to probe the converse piezoelectric response of the ferroelectric thin film. By establishing a stable electrical contact in the repulsive regime, an alternating current (AC) excitation, $V_{AC}\cos(\omega t)$, is biased between the tip and a grounded bottom electrode. The resulting localized electric field induces a respective mechanical deformation of the lattice via the converse piezoelectric effect, which results in a cantilever displacement (D) characterized by both amplitude (A) and phase ($\varphi$). As depicted in Ext. data Figure 8, the directional coupling between the applied field and the spontaneous polarization vector (P) dictates the electromechanical contrast: a parallel orientation results in an in-phase expansion (contraction), while an anti-parallel orientation yields a 180° phase shift. This displacement is resolved by a four-quadrant position-sensitive photodetector (PSPD), where the vertical deflection (A-B) and lateral torsion (C-D) signals are decoupled to isolate the longitudinal $d_{33}$ and shear $d_{15}$ contributions, respectively. Through the integration of high-bandwidth lock-in amplifiers, the vectorial components of the piezoresponse are mapped with nanoscale spatial resolution. The forces acting on the cantilever are given by the force = $k * \delta$, where k is the spring constant calculated using solid mechanics equations for its shape, dimensions, and material properties.[40] The deflection is constantly monitored using laser feedback, and the forces and sensitivity are calibrated using a force vs. distance curve for a particular sample. The spring constant of the cantilever is calibrated using Sader's method.[41]

The following equations govern the response of a PFM. [42]

$$D=d_{33}V \quad -(1)$$

$$D(t)=d_{33}V_{DC}+d_{33}V_{AC}\cos(\omega t) \quad -(2)$$

$$V = V_{DC} + V_{AC}\cos(\omega t \pm \Phi) \quad -(3)$$

$$\text{Piezo Response} = R_r = A_r * \cos(\varphi_r) \text{ (suffix r could be x, y, or z)} \text{—}(4)$$

$$A = A_0 + A_1\cos(\omega t+\varphi) \quad -(5)$$

$$\theta = \tan^{-1}(\varphi_x / \varphi_z) \quad -(6)$$

The piezoelectric sample response to the applied voltage is given by Eqn 1, where $d_{33}$ is the piezo coefficient, $V$ is the applied voltage, and $D$ is the displacement of the cantilever derived from the electrical signals of the detectors. Equation (2) shows the relation between the effective displacement and the net electrical signal (as shown in equation 3) applied across the cantilever and the sample. The measurements were performed in both off-resonance and in-contact resonance modes. The actual piezo response of the sample is given in Eq. 4, where $A$ is the amplitude of cantilever deflection, including sample response and the cosine of the phase difference. The amplitude A could be further resolved as shown in Eq. 5, where $A_0$ is the cantilever deflection and $A_1$ is the sample response, along with the respective phase change.

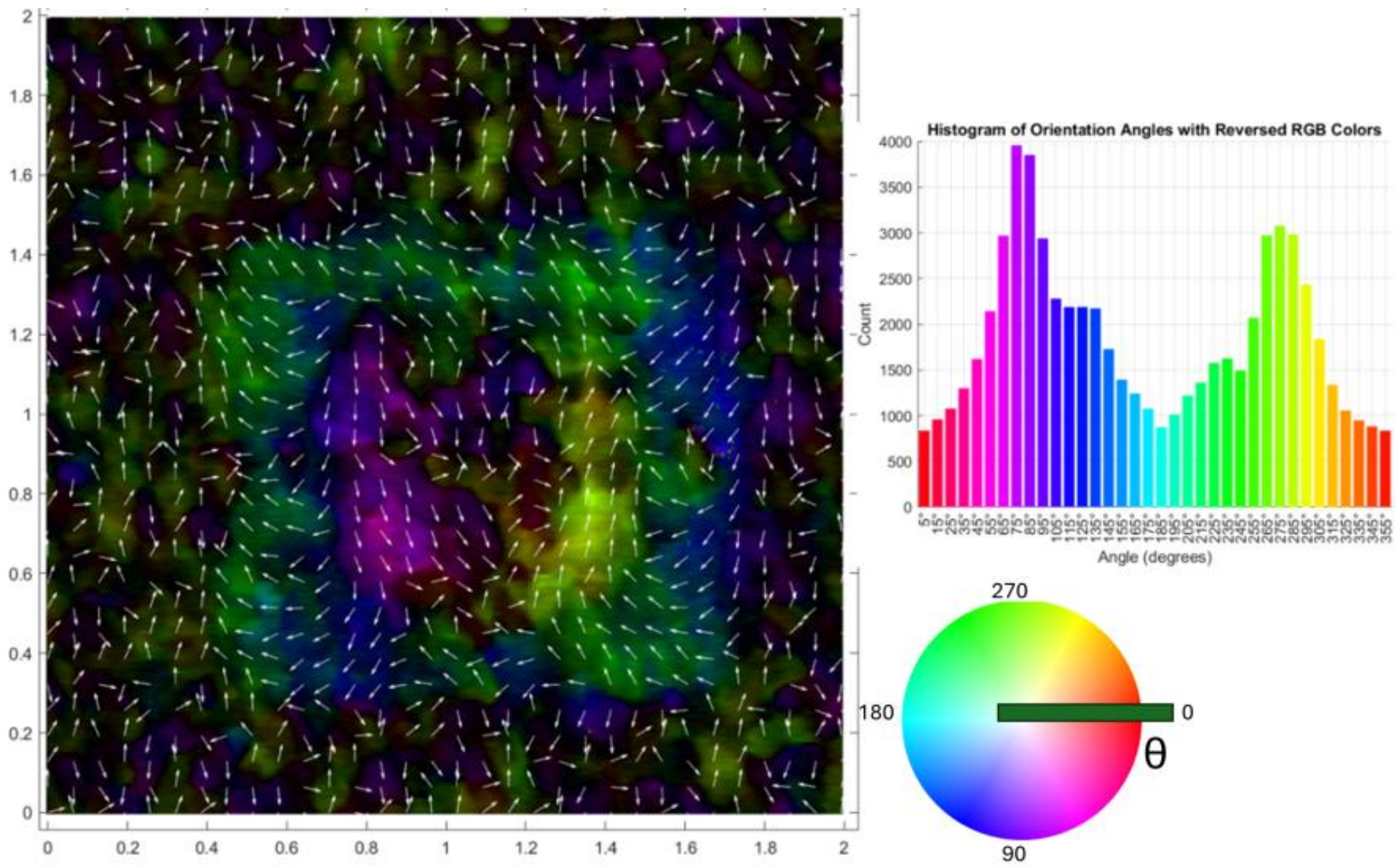

***Figure 9** shows the colour map with overlaid arrows representing the polarisation direction on the surface of the thin film poled area.*

The analytical framework for 3D polarization mapping, as outlined in this study, treats the discrete vertical and lateral piezoresponse as orthogonal vectorial components. By implementing a pixel-by-pixel correlation algorithm, we can reconstruct the local electromechanical signals into a two-dimensional vector field, providing a high-fidelity map of the in-plane and out-of-plane polarization orientations. Furthermore, by incorporating datasets acquired from orthogonally rotated scans, we achieve a full three-dimensional reconstruction. This method yields a unique, high-resolution directional representation of domain architecture that serves as a powerful tabletop alternative to computationally intensive simulations or destructive transmission electron microscopy (TEM) to obtain similar information.

The disparity between conventional 2D phase imaging and our 3D reconstruction is evidenced in the analysis of the poled PZT region shown in Figure 9. While the isolated vertical and lateral phase profiles offer a fragmented view of the ferroelectric state, the combined resultant vector calculations reveal the real domain morphology, overcoming the inherent projection limitations of the cantilever geometry. This approach facilitates the deterministic mapping of domain distributions and the predictive modeling of phase contrast across complex polycrystalline surfaces.

Since ferroelectric domains are polar and possess a unique polar axis, it results in unique domain wall angles between them, which facilitates an identification route[43,44]. Such identification is possible via reconstruction of the data upon calculating the actual orientations of the domains with respect to the sample normal by using the grain orientation relation with the crystallographic symmetry.[45] Since each crystal system possesses a signature domain structure, by reverse modeling the piezo-response and reconstructing them with real polarization angles, we could correlate with the plausible crystalline phases present in the sample.[46] The interrelation between the polarization direction, with respect to the sample normal, which relates to the respective crystallographic (*hkl*) grain orientation for tetragonal, orthorhombic, and rhombohedral, is provided in the supplementary (S6).

### 7.3. Crystallography and Electromechanical Correlations

While X-ray diffraction (XRD) remains a standard method for identifying the constituent phases and preferred orientations in polycrystalline thin films, its nature as an averaging technique inherently obscures the spatial heterogeneity of the material. In systems near a morphotropic phase boundary (MPB), films typically manifest as a complex mosaic of grains, each possessing distinct crystallographic symmetries such as tetragonal (T), rhombohedral (R), or orthorhombic (O) and unique domain architectures. Consequently, the local electromechanical response is governed by the specific orientation of a grain and its interaction with the surrounding elastic environment.[47]

Bridging this gap requires moving beyond statistical distributions to resolve how a specific (111)-oriented rhombohedral grain is functionally influenced by a neighboring (110) orthorhombic grain.[48] By establishing the geometric projections of the spontaneous polarization vector onto the measurement axes, one can analytically predict the characteristic piezoresponse signatures. For example, a (001)-oriented tetragonal grain exhibits a purely longitudinal response, whereas a (110) orthorhombic grain generates a hybridized signal, characterized by high vertical sensitivity and asymmetric lateral response along a single planar axis (refer to supplementary figure S6).[49]

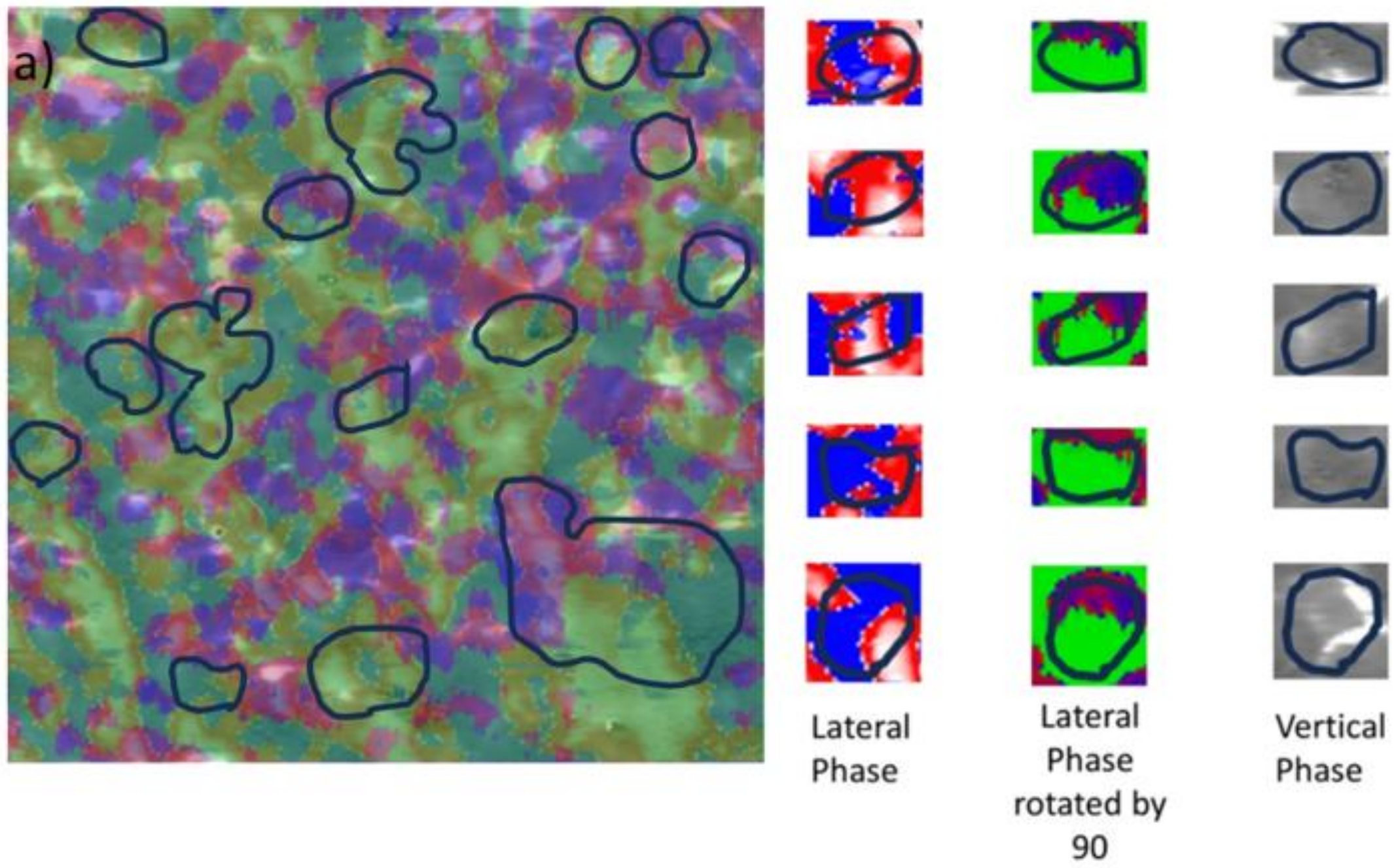

***Figure 10.*** *PFM phase data of BCZT thin film overlaid together with vertical, lateral, and 90° rotated lateral phases.*

Such spatial resolution is fundamental to domain engineering, where the objective is to precisely modulate material properties via external stimuli, including electrical bias, epitaxial strain, chemical doping, and thermal gradients. Synthesizing local phase identification with 3D piezoresponse mapping thus provides the necessary framework for the deterministic control of domains in high-performance ferroelectric applications. Figure 10 provides a complete visualization of PFM data acquired from the BCZT thin film. There are three distinct phase channels overlaid for better understanding: the lateral phase at 0° (represented by the red/blue color map), the orthogonally rotated lateral phase (green/purple), and the vertical phase at 0 (grayscale). Detailed analysis of the highlighted grains reveals a critical difference between phases and amplitude. While the phase clearly delineates the domain boundaries, amplitude is necessary to quantify the signal strength. To overcome the limitations of 2D imaging, we have attempted to develop a pixel-by-pixel correlation algorithm to reconstruct 3D vector maps of polarization orientation. Integrating vertical and lateral phases (at 0° and 90° rotations) facilitates identification of specific grain orientations (Table 1).

Table: 1; shows the piezo response of particular grain orientations for a given Crystal system.

| **Vertical (V)** | **Lateral *ϕ=0* (L0)** | **Lateral *ϕ=90* (L90)** | **Plausible orientation of a phase** | **Inference** |
|---|---|---|---|---|
| **1** | **0** | **0** | **(001)Tetragonal** | Purely out-of-plane polarization. Unless we are sure about symmetry we Could not distinguish between (001) Tetragonal *or* (111) Rhombohedral. |
| **0** | **1** | **0** | **(010) Tetragonal** | Purely in-plane polarization (b-domain) aligned with "0" scan. |
| **0** | **0** | **1** | **(100) Tetragonal** | Purely in-plane polarization (a-domain) aligned with "90" scan. |
| **1** | **1** | **0** | **(110) Tetragonal** | Tilted polarization (component in-plane along "0" scan). |
| **1** | **0** | **1** | **(101) Tetragonal** | Tilted polarization (component in-plane along "90" scan). |
| **0** | **1** | **1** | **(011) Orthorhombic** | Tilted polarization (component of both in-plane scan). |

While binary logic identifies the presence of polarization components, it lacks the resolution to differentiate between primary high-symmetry poles and proximal low-angle planes. High-fidelity visualization is only achievable by integrating crystallographic structural data, which allows for the derivation of deterministic response ratios across different PFM scan planes.

This approach enables the precise mapping of angular relationships and the identification of distorted lattice regions that would otherwise be wrongly classified.

.

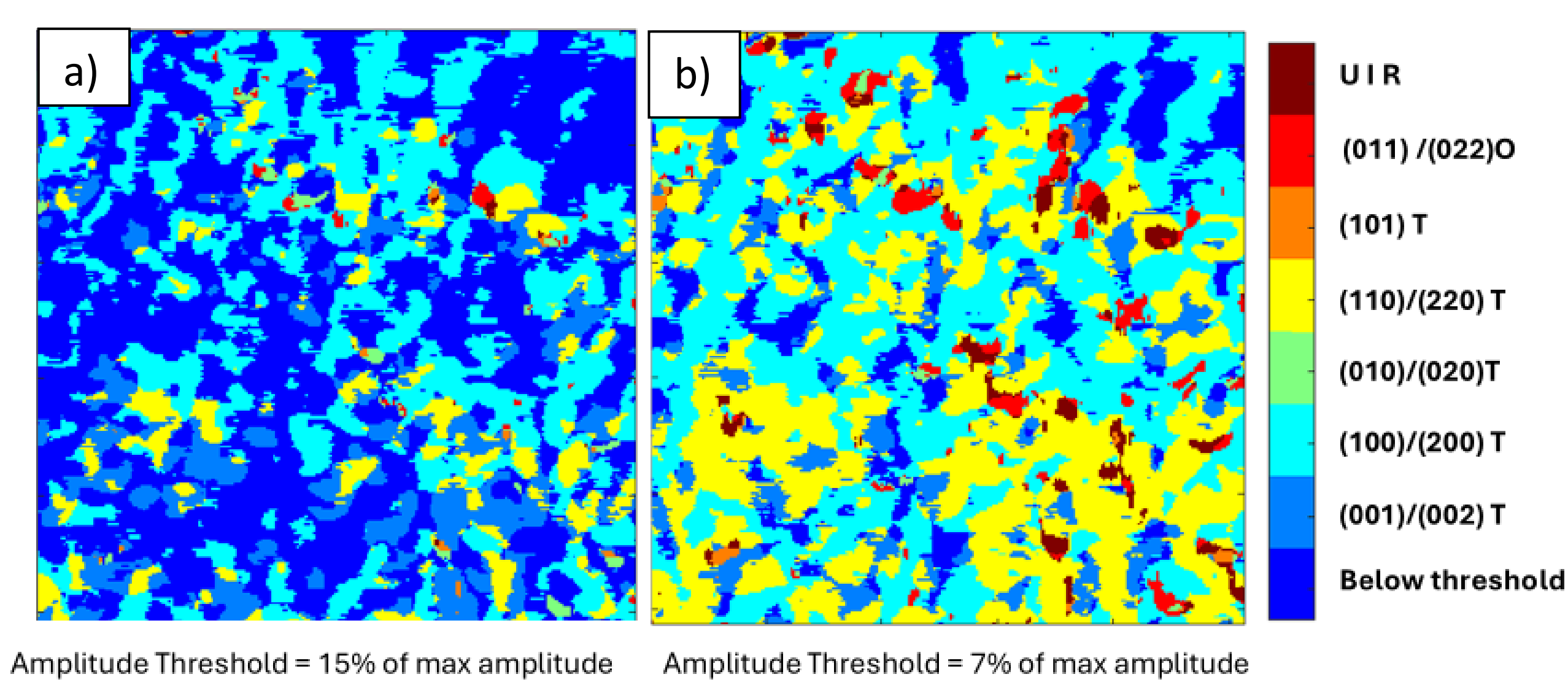

***Figure 11**: Possible grain orientation reconstructed by different planar responses of PFM. a) and b) show the same with an amplitude threshold 15%, and b shows 7%.*

Figure 11 (a) shows the possible grain orientation predicted using Table 1 with a <15% amplitude threshold given as zero. And b) shows the same with 7% threshold. The details of the amplitude threshold are detailed in the subsequent section. Any data point below the threshold is noted by lower threshold regions (LTR). By integrating the indexed response with known material phase information, grain orientation can be predicted based on the magnitude of the PFM signal. When a signal falls below the specific threshold defined by the instrument parameters, it is indexed as a below-threshold value. The details of the relation between cantilever orientation and the respective polarization variants of the heterophase systems are detailed in S6 of the supplement. However, there are instances where the data sets are unidentifiable within the plausible heterophases (orthorhombic, tetragonal, and rhombohedral in the case of BCZT) and grain orientations present, but still possess strong PFM signals. To avoid the misclassification of these ambiguous signals, these zones are categorized as Unidentified Regions (UIR) rather than being assigned a definitive crystallographic index.

### 7.4. Validation of the piezoresponse signal threshold

The electrical signal generated from the deflection of a cantilever upon applying a bias through the tip possesses a physical limit that comes with the hardware of the machine, based on the resolution it could capture. The min amount of signal that can be detected by the PSPD varies between μV to a few V. When the amplitude response is taken into the binary algorithm, the ambiguity of the true signal and the noise of the instrument needs to be cross-verified. This determines the zero signal of the PSPD. In order to differentiate the noise from the zero signal scenario of a sample, a nonferroelectric material like metal (Pt/Si substrate) was used to determine the instrumental zero signal condition with the same probe. Later, the obtained l amplitude data in the image form is provided as an input to the algorithm, which calculates the ratio between the maximum and minimum amplitude signals of the PSPD. The amplitude ratios of Pt/Si samples were as low as ~1%. This confirms that the zero signal from the sample could be as close to 1% of the background electronic noise. Hence, in this study, to obtain information about the strength of the various components and to correlate it to the crystallographic orientation and its relation to the polar axis, it is important to finalize this threshold ratio of the maximum and the zero signal arising from the ferroelectric samples and fix it at ~7%. Though the calculated minimum could be as low as ~1%, in order to take into account the slip and slide of the cantilever due to morphology, which adds up to weak signals, the threshold was kept several times higher than the calculated low threshold. Increasing the threshold to relatively higher values (~15%) could suppress the information from the sample, as shown in Figure 11 (a). The same data, when analyzed using a threshold value of 7%, provides the difference between two crystallographic orientations that could have a close angular relation with the polar axis of the given material. Figure 11(b) shows the analysis of the same data as 5(a) with 7% threshold ratio, and it could distinguish between (100)&(110) Tetragonal phase with a small fraction of (011) O phase. Throughout this study, the threshold ratio was fixed at ~7%. Complementary studies (TKD from TEM) could facilitate further reduction of these threshold values.

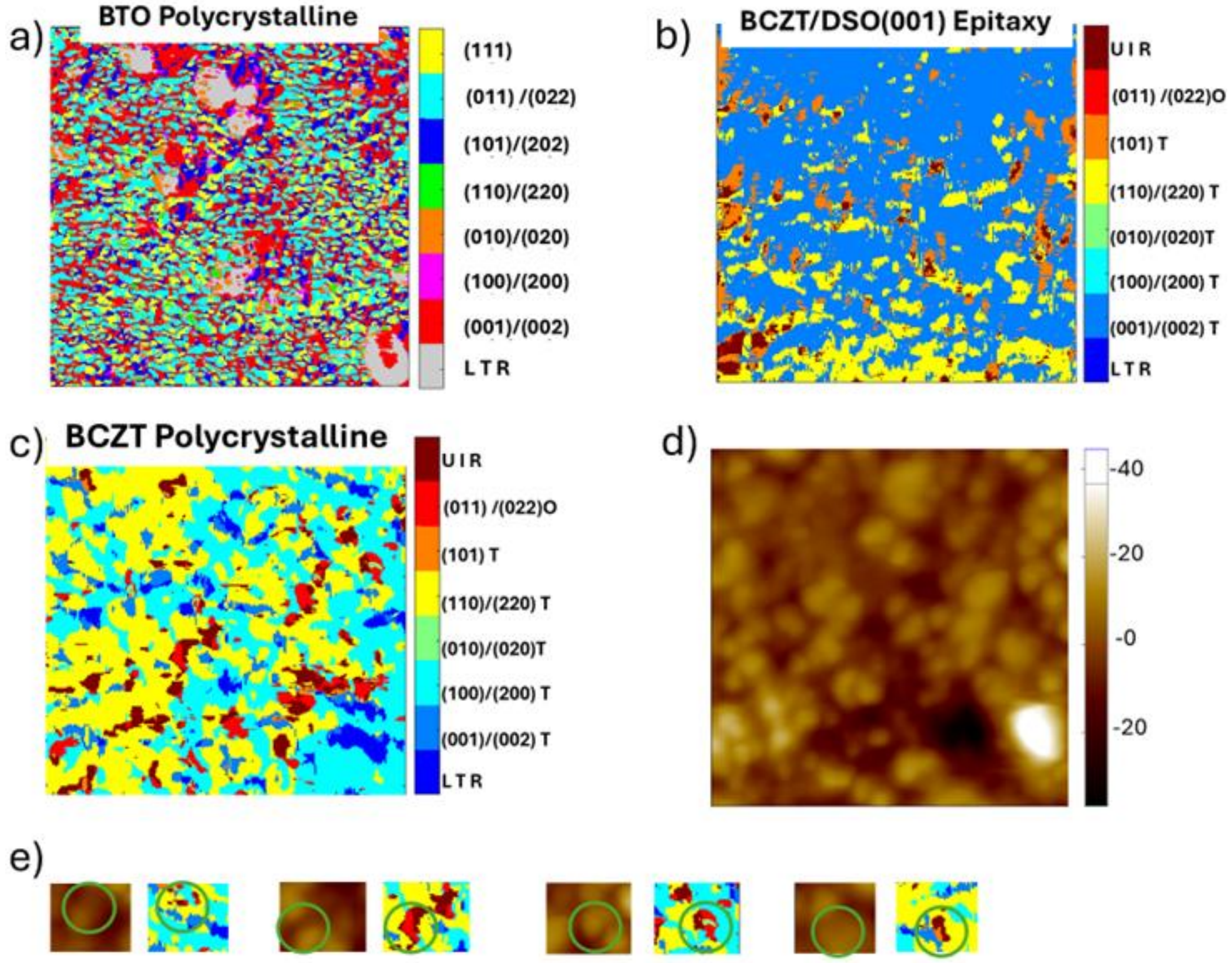

***Figure 12*** *a) shows the possible grain orientation of BTO film b) shows the orientation map of the BCZT (001) / DyScO₃(001) epitaxy and c) shows the grain orientation map of polycrystalline BCZT film and f) shows the 3D phase resultant map d) morphology of BCZT polycrystalline thin film, e) shows the comparison of morphology and unidentified regions(UIR).*

To verify the outcome of the analysis, the studies were carried out on a homophased polycrystalline $BaTiO_3$ film with the same growth conditions as the BCZT thin film. Figure 12 (a) shows the amplitude-based orientation map; most of the area is in agreement with the crystallographic relation. Then similar analyses were carried out in the epitaxy of BCZT (001) grown on $DyScO_3$ (001), shown in Figure 12 (b). The epitaxial thin film was observed to possess a mixture of tetragonal and orthorhombic phases, as previously observed.[50] Using the binary amplitude algorithm developed, the predicted orientations primarily consist of (001) and (010), with a lower contrast observed for (011) or (022) planes. This is likely due to local strain gradients or grain boundaries where the lattice is distorted into closer low-angle neighbors, causing the piezo-response to be "off-axis" from the primary phases. The similar type of response was seen in polycrystalline BCZT, also shown in Figure 12(c) and (d) shows the morphology of the polycrystalline thin film. The algorithm predicted that the entire area was dominated by (110)T and (100)T types with a minimum of (011)O signature. The surface also presents some of the area of titled response, which cannot be determined to be either tetragonal or orthorhombic, or it is distorted in between these two phases. But those present only at the interfaces between the orthorhombic and tetragonal predicted orientations, which is a signature most commonly seen in the parent compound of BCZT–BTO. The observation leads to evidence of the origin of a new polarization invariant plausibly monoclinic type distortion due to the bending stresses suffered by the BCZT sample, and similar observations are seen in literature and also supported by Raman studies of BCZT thin films.

It is important to note that it was later confirmed through Raman studies that irreversible transformation to monoclinic could be present due to the bending stresses applied to the sample. Thus, the analysis technique proves to be an effective tool to detect polarization invariants originating in heterophased polycrystalline systems across MPB regions undergoing changes under different conditions.

### 7.5.Three-Point Bending:

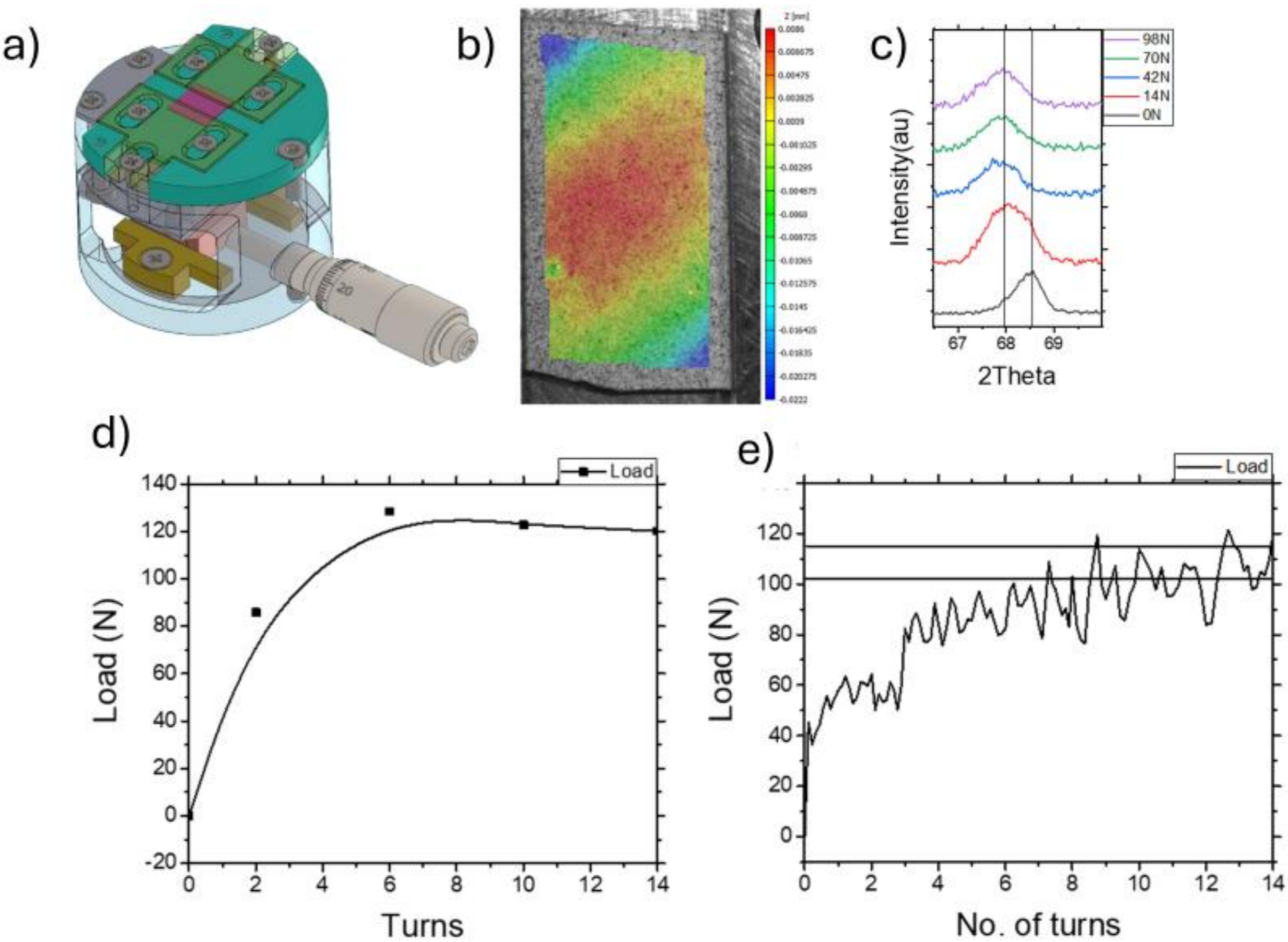

***Figure 13*** *a) shows the configuration of the sample mount, b) shows the stage mounted on a stage along with si substrate coated white pain and speckle patterns shows the z depth from the camera to sample center. C) shows the GI XRD of Si(004) peak from the bending stage. d) Shows load vs no of turns from XRD experimentation (e) shows load vs no of turns from DIC experimentation.*

Figure 13 (a) shows the stage with all the parts in an assembled construction. The bottom wedge will be fitted in the slot and then fed through an Allen screw with the same pitch as a screw gauge as shown in Figure 13. With a pin loaded with a spring, it will translate the horizontal displacement into vertical displacement. The top plates will act as support for the sample to be held on top using screws on extended circular plates.

### 7.5.1. Standardization of the bending Stage: 3D-DIC & GIXRD

To calibrate the custom-designed stage fabricated by APE Research (Italy), 3D Digital Image Correlation (3D-DIC) was employed to map the mechanical response of Si(100) substrates (E = 130 GPa}). By tracking a speckle pattern with dual high-resolution cameras at 10 fps, the system recorded a maximum out-of-plane displacement of 0.1 mm over 14 micrometer screw turns. This corresponds to a calculated load of approximately 100 N per 360° rotation and a strain value of 0.0375, which remains safely below the silicon fracture point. The resulting strain gradient follows simply supported beam theory, inducing in-plane tension at the top surface and in-plane compression at the bottom. These DIC-derived parameters were further validated through in situ X-ray diffraction (XRD) using a Bruker D8, ensuring a precise correlation between screw actuation and the physical stress applied to the deposited BCZT films.

Then, the displacement of Z along the normal was observed using the 3d DIC technique, and the displacement for the total of 14 turns in the deflection of the Si substrate was observed to be approximately 0.1 mm.

For the deflection of 0.1 mm and Si (100) having a young's modulus of 130 GPa,

$$P = \frac{48EI\delta}{L^3} \qquad \text{–(7)}^{51}$$

$\delta$ = deflection in mm, $E$ = Young's modulus, I = moment of inertia, $L$ = support span, mm

To calculate the respective load applied to the Si (100) substrate (E = 130 GPa), analytical modeling was correlated with experimental 3D-DIC and in situ XRD data. For a deflection of 0.1 mm over an 8 x 8 mm area, 3D-DIC measured a strain of 0.0375, corresponding to a total load of 101 N across 14 micrometer screw turns (~7 N per turn). Interestingly, XRD peak analysis revealed a shift toward smaller 2 theta values, indicating an anomalous out-of-plane tensile strain. While isotropic materials typically exhibit out-of-plane compression under surface stretching, this behavior is attributed to the elastic anisotropy of the single-crystal silicon. Quantitative fitting of the XRD peaks yielded a calculated load of 122 N (~8.5 N per turn). The macroscopic DIC results and high-precision XRD measurements show strong agreement in the saturation load, as shown in Figure 13(d and e). The wedge-driven, spring-loaded mechanism of the portable stage thus provides a consistent and reproducible loading profile, confirmed by the convergent results of both calibration techniques. The analytical values calculated from the bending state were correlated with Grazing Incidence X-ray Diffraction (GI-XRD) measurements, where lattice strain was independently verified via the shift in the 2theta positions of the (110) and (111) Bragg reflections.

### 7.5.2. Contact pressure measurement:

#### 7.5.2.1.Mechanical Characterization and Stress Analysis

The maximum flexural stress at the midpoint of the tensile surface was calculated using the standard Euler-Bernoulli beam theory:

$$\sigma = \frac{3FL}{2wh^2} \qquad \text{–(11)}^{51}$$

Under the specified loading conditions, the maximum flexural stress was determined to be approximately 485.95 MPa.

$$P = \frac{F}{A} \qquad \text{–(12)}^{51}$$

This yielded a localized contact pressure of 2.36 GPa.

The calculated contact pressure of the sample holder that applies directional 3-point bending to a thin film on a substrate indicates a value of 2.36 GPa. In this study, the pressure is applied through 3-point bending, in which the sample is already clamped with a Pt/Si substrate. Consequently, the current 98 N load is roughly half the magnitude required to observe the pressure-induced ferroelectric-to-paraelectric transition identified in the bulk study but is enough to change the already strained thin film phase fractions[52]. Hence, it is evident from the Raman studies that it could be the same distorted monoclinic phase that emerges in the bulk ceramics that appears under bending stresses of BCZT thin films.

### 7.6. Force vs. Distance with Fixed-Point Measurements and Mapping:

To validate the coexistence of tetragonal (T) and orthorhombic (O) phases typical of MPB materials like BCZT, we employed FD pinpoint mapping. Because BCZT's mechanical properties (stiffness, Young's modulus) are sensitive to bond strength and crystal structure, different phases exhibit distinct "mechanical signatures."

Generally, the physical property of these phases of the same material depends on density, bond strength, and crystal structure. However, in materials like BCZT, the presence of five atoms with varying valencies means that phase stabilization is composition-dependent. These chemical variations alter bond strengths and covalency, leading to a drastic impact on mechanical properties like stiffness and Young's modulus. In this work, we also propose that we could characterize these heterophased regions that have different phases, with varying stiffness, by performing F vs. d pinpoint mapping. The studies reveal similar gains, with a 20 to 70 GPa variance compared to other grains, indicating the presence of heterophases in a given grain. While the film may have different densities depending on fabrication parameters, achieving an exact match with bulk values is not strictly necessary for this analysis; however, the considerable amount of contrast in the values observed is a strong indicator of regions with different mechanical behavior (mainly stiffnesses). Using a 120 nN load and a tip with a 3 N/m spring constant ensures sufficient force to identify deformation contrast between grains without damaging the sample. Consequently, these grains are identified as distinct phases where deformation is more pronounced, and the modulus deviates by more than one-third from the tetragonal phase. This contrast provides valuable insight into materials where the film has a history of mechanical bending.

## Acknowledgement

The authors would like to thank DST-SERB India for the funding (Grant No.: DST/SERB/EMR/2017/003159/MMM). SATHI-CISCoM center at IIT Hyderabad, funded by DST India, SR/SATHI/2022/247. We thank Prof. Ramji M., IIT Hyderabad, for the DIC support studies in the calibration of the 3-point bending stage. We also acknowledge 'APER' APE Research (Italy) for their support in the design and precision fabrication of the in-situ three-point bending stage. We would like to thank Prof. Saswata Bhattacharyya, IITH, for the technical discussions.

## Data Availability Statement

The data and the MATLAB code will be provided on request.

## Conflict of Interest

The authors declare that there is no conflict of interest.

**Supplementary Material**

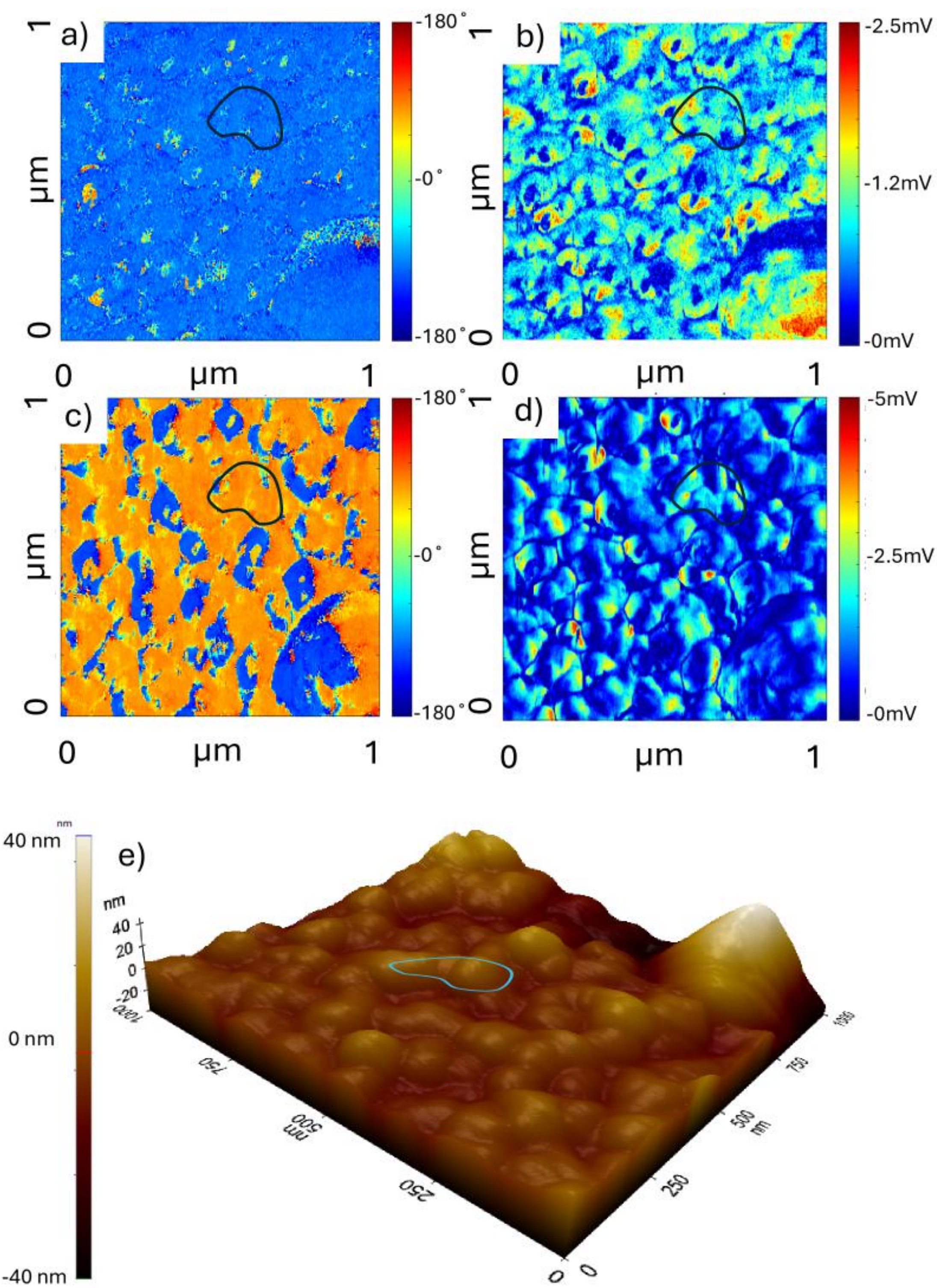

*S 1* ***a,b.*** *PFM out of plane phase and amplitude,* ***c,d.*** *PFM lateral phase and amplitude and* ***e.*** *Topography of BCZT thin film.*

Figure S1 characterizes the standard PFM response alongside the physical topography of the BCZT film. Panels (a) and (b) represent the out-of-plane PFM phase and amplitude images respectively, which highlights the switching characteristics of thedominant vertical components . Panels (c) and (d) show the lateral (in-plane) phase and amplitude, which reveal the torsional movement of the AFM cantilever. Figure S1 (e) shows the 3D topographic map showing the surface roughness and grain structure of the film. The highlighted region shows the domain pattern extending across grains crossing the grain boundary.

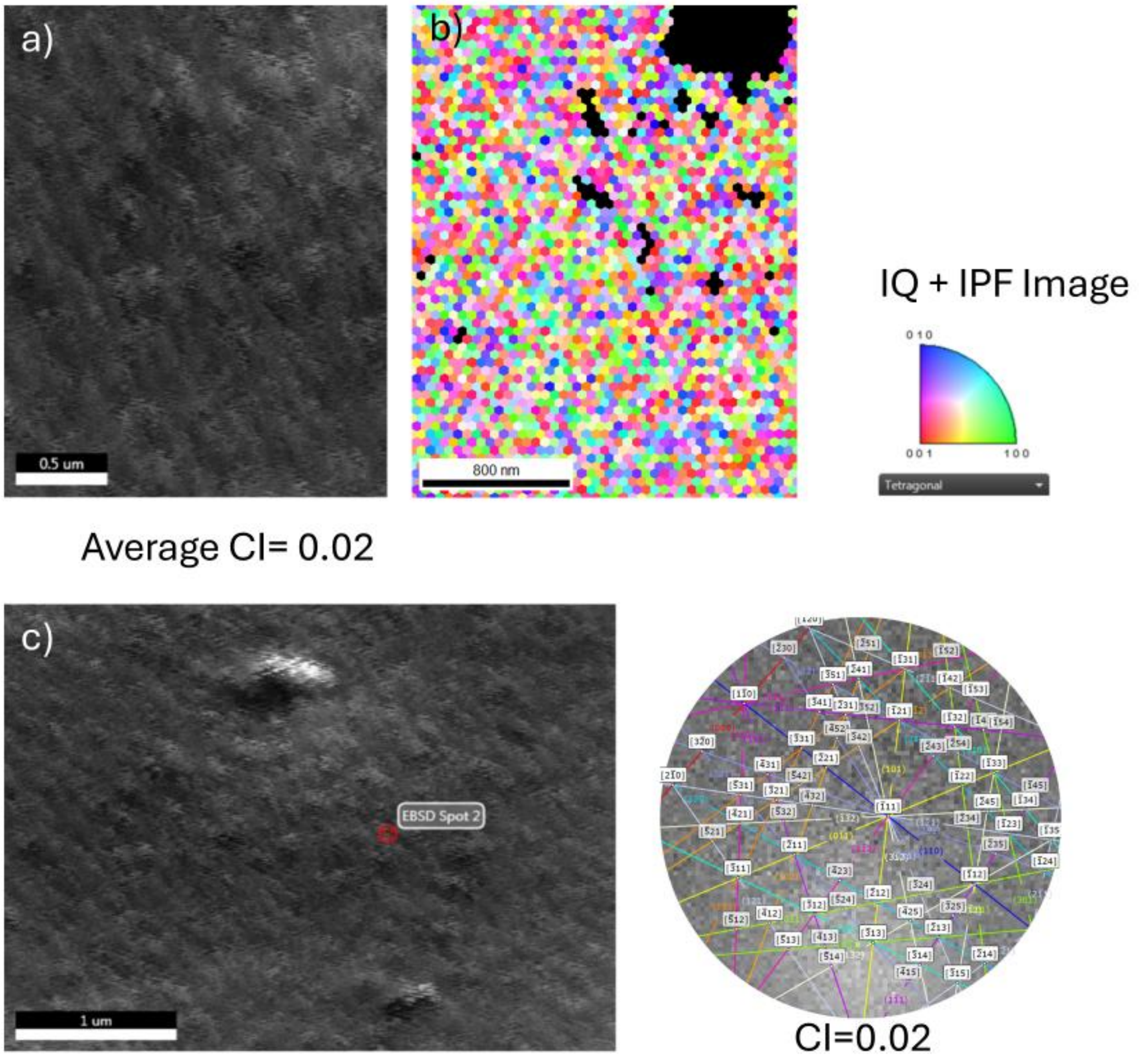

*S 2 a) shows the SEM image of EBSD of BCZT thin film, b) shows the indexed orientation image of EBSD, and c) shows the Kikuchi pattern taken in a specific point.*

Figure S2 focuses on the structural and crystallographic orientation of the BCZT thin film using Electron Backscatter Diffraction (EBSD). Panel (a) is a standard Scanning Electron Microscope (SEM) image of the surface, while (b) is an Inverse Pole Figure (IPF) map that colors grains based on their crystallographic orientation. Panel (c) displays a Kikuchi pattern captured at a specific spot, which is the raw diffraction data used to index the crystal lattice. The average Confidence Index (CI) of 0.02 is noted, indicating the ambiguous reliability of the indexing across the film due to  relatively lower overlaying signals of multiple and tiny grains (~ 20-50nm).

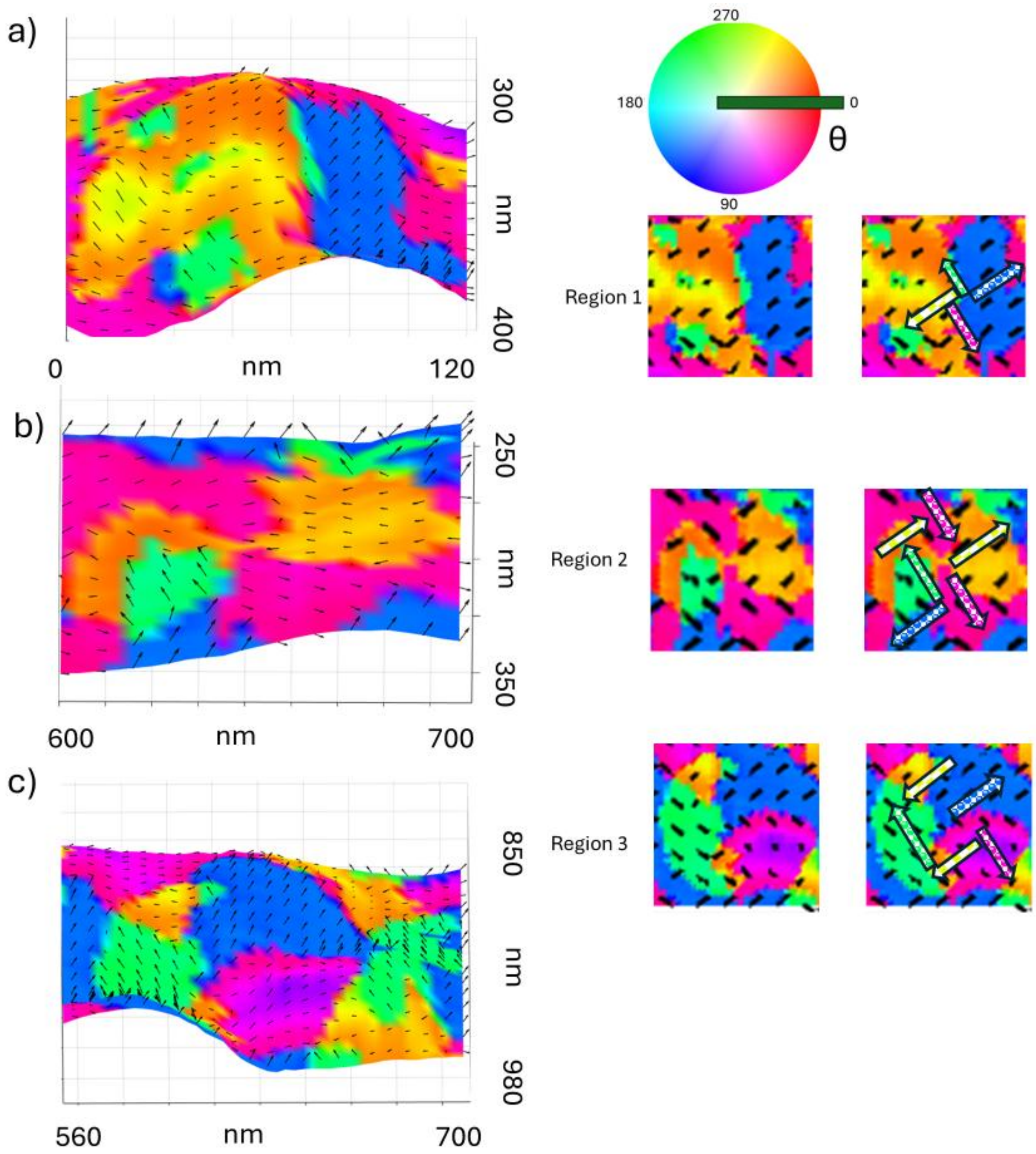

*S 3 a),b) and c) shows the 3D overlay of morphology and 3D resultant polarization components of region 1,2 and 3.(right) shows the 2D view of the same regions.*

Figure S3 provides a detailed 3D overlay of the surface morphology with resultant polarization components for three specific areas, labeled as Region 1, 2, and 3 chosen from the morphology presented in the main text (Fig 1). The black vectors superimposed on the topography indicate the local direction of the polarization in 3D space. The right column of images, indicate the top view of the same regions with simplified arrow diagrams illustrating the internal domain arrangements at the respective regions. Such analysis facilitates visualization of the alignment of polarization vectors along with the physical morphology of a given microstructure.

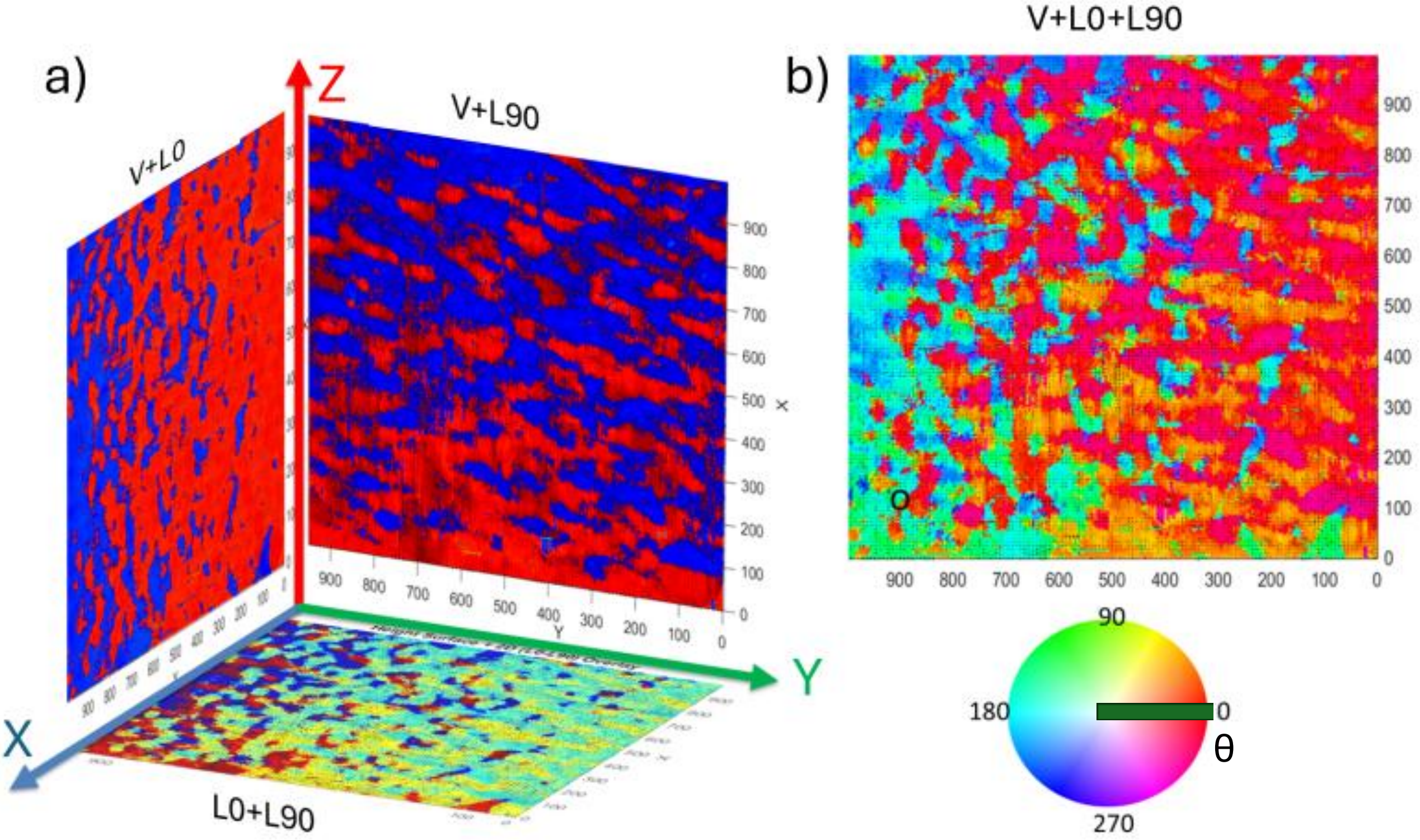

*S 4 a) shows the reconstructed 2D polarization vector maps across the **XZ, YZ, and XY** planes and b) shows the 3D resultant PFM of BCZT(001) epitaxial thin film of 80nm over $DyScO_3$ (001) single crystal substrate, a shows the polarisation map created to understand the domains better.*

Figure S1 shows the Piezoresponse Force Microscope (PFM) image for the 80 nm epitaxial thin film of BCZT (001) grown on a $DyScO_3$(001) single crystal substrate. It includes a 3D visualization of the vertical and lateral PFM signals overlaid on morphology (underneath) with reference to respective pixels, to show the domain structure obtained from the projections in multiple planes. The polarization map (b) is specifically constructed to provide a comprehensive understanding of the domain orientations in the scanned region. The color-coded azimuth wheel indicates the specific direction of the 3D resultant polarization components with respect to the long axis of the cantilever, indicated as a rectangular bar in the color wheel.

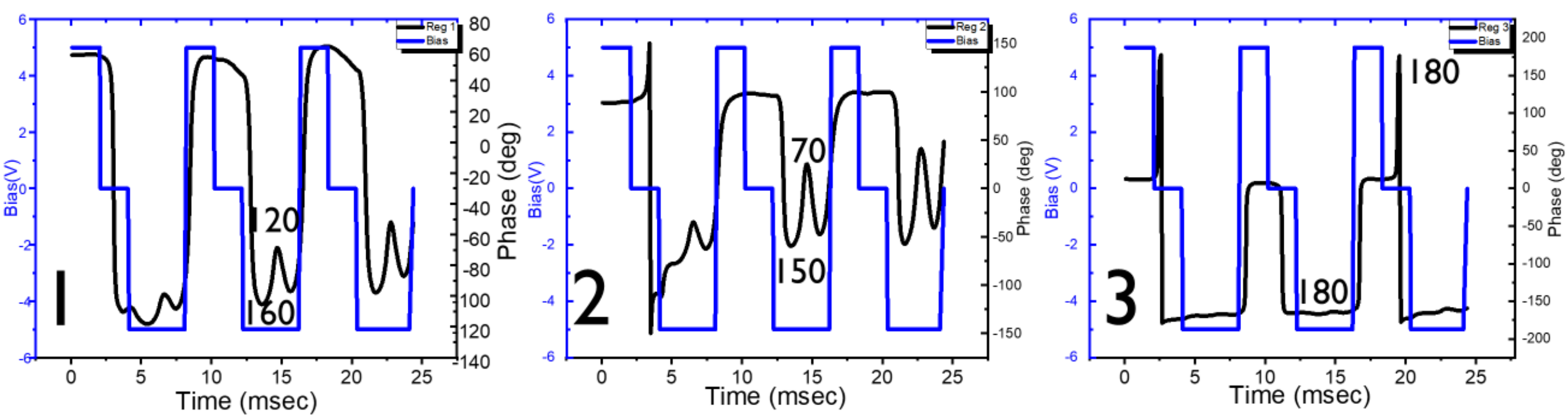

*S 5 shows time dependant switching of phase component of specific regions.*

Figure S4 illustrates the dynamic behaviour of the ferroelectric domains under an applied bias over time. The graphs show the phase component switching as a function of voltage captured over a time period in three different regions BCZT polycrystalline thin films. The square blue waves represent the bias applied, while the black curves track the resultant phase shift recorded by the inbuilt lock-in amplifier. Notable phase angles such as 180˚ and 70˚ are marked to indicate the degree of domain reorientation occurring in each specific region shown in main manuscript Fig 2. The region 1 & 2 have intermediate states, whereas In Region 3(ferroelastic), which is dominated by a single 90º ferroelastic configuration, the switching becomes a competition of directions results in the clean 180º phase jumps.

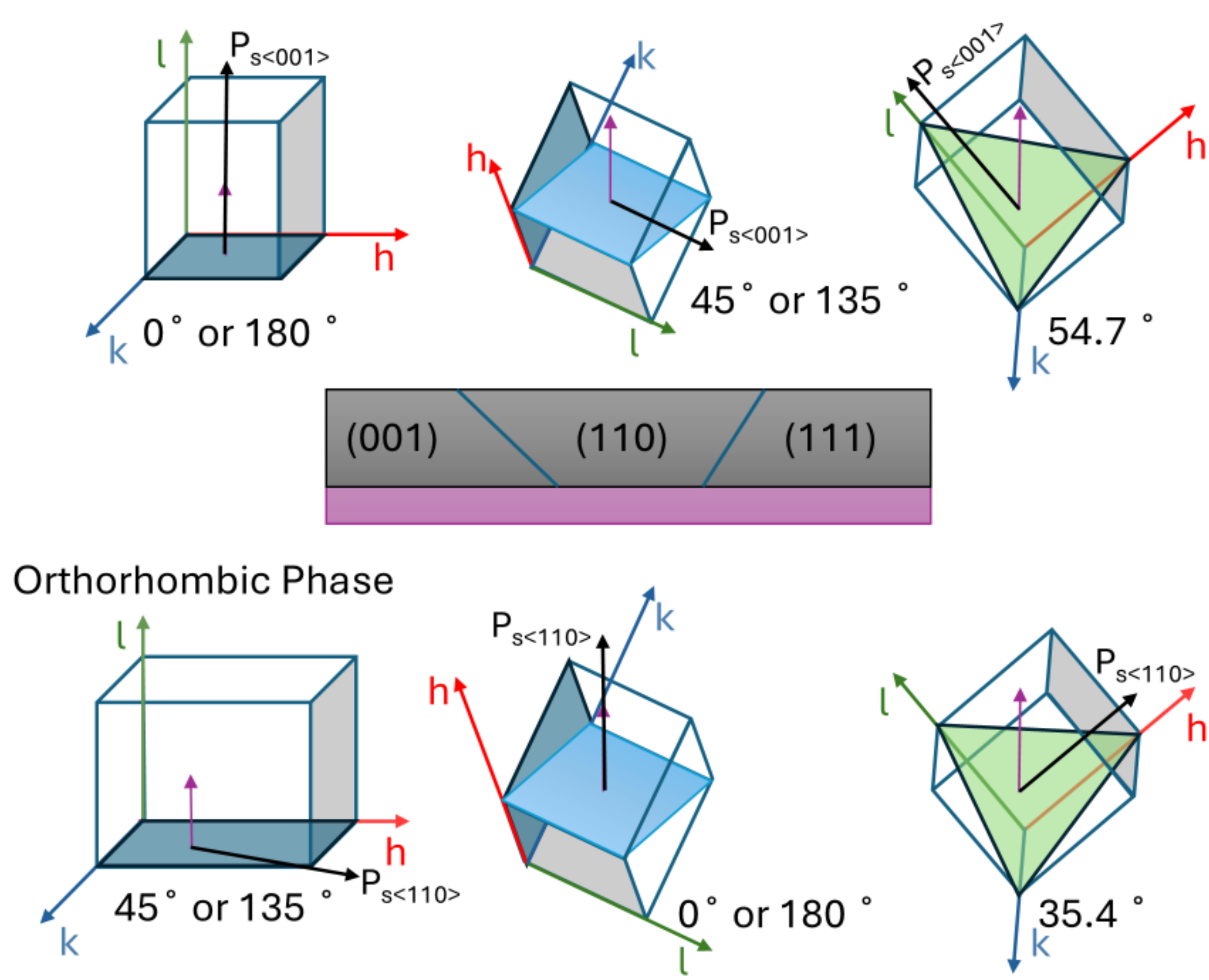

*S6. Shows Tetragonal and Orthorhombic crystal structure present in BCZT and the angular relation between the polar axis and the grain orientations.*

Figure S 6 shows the orthorhombic and tetragonal crystal structures and the angular relation between their respective grain orientations of these phases and the polarization vector direction. Utilizing these angular relation in cartesian co-ordinates and deconvoluting it in three different axis and their respective quadrants, facilitates us to identify the strengths of the components and which subsequently indicates the crystallographic orientation of a given ferroelectric phase. This technique facilitates identifying the unique signatures of all possible grain orientations which are present in the device and is known by independent XRD studies. With the knowledge of prior studies about the possible grain orientations and combining it with the deconvoluted PFM signal a polarization-microstructure correlation could be made. The table given below details the principle behind the analysis for the chosen grain orientations of the tetragonal phase.

Table 1: Resultant response *process in the table for some of the grain orientations and the respective amplitude response in the Ax-L0, Ay-L90 and Az- V .*

| ***R*** | $A_x$ | $A_y$ | $A_z$ |
|---|---|---|---|
| (100)T | >0 | 0 | 0 |
| (110)T | >0 | >0 | 0 |
| (111)T | >0 | >0 | >0 |
| (001)T | 0 | 0 | >0 |

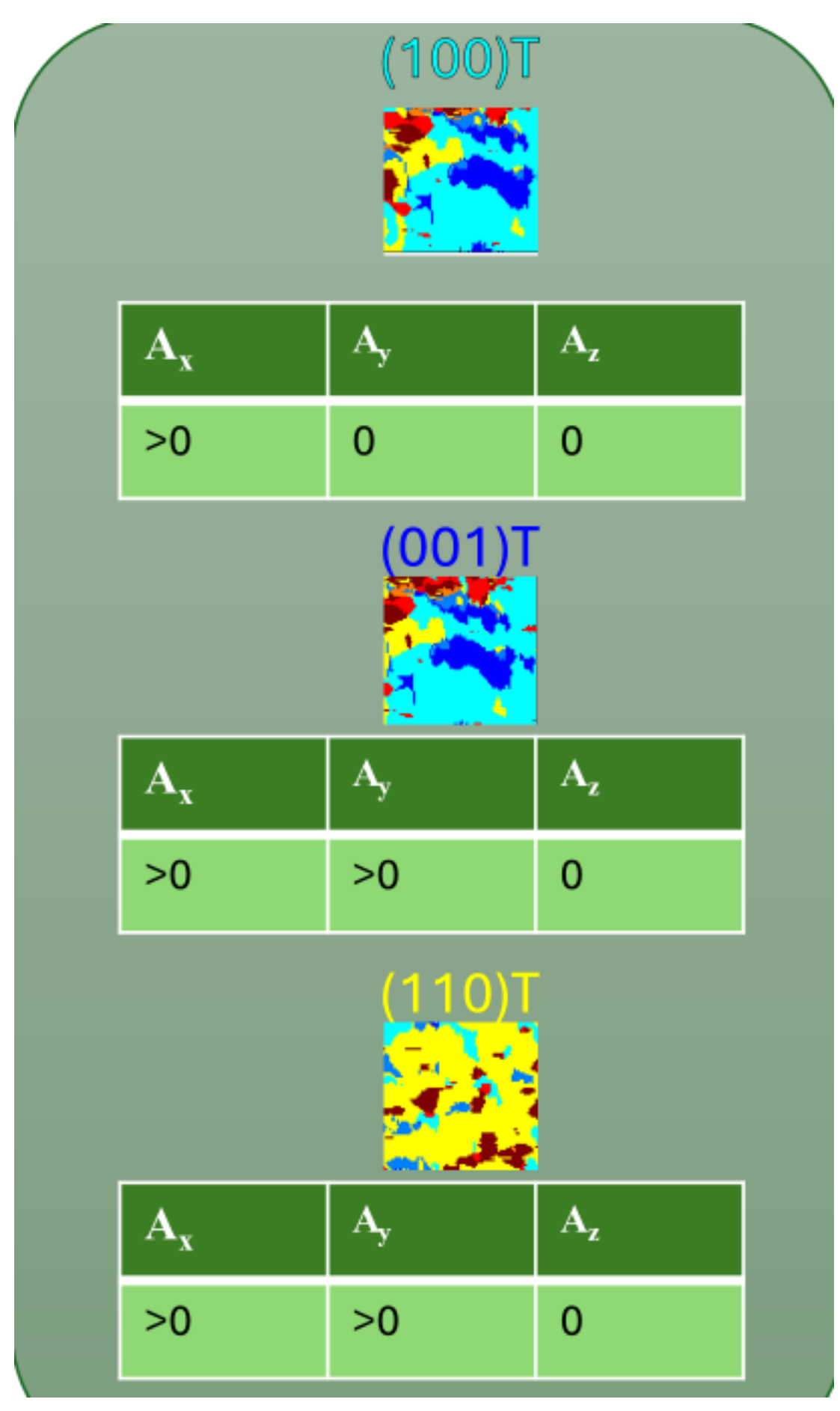

*S 7. Shows the process in the table for some of the grain orientations and the respective amplitude response in the Ax-L0, Ay-L90 and Az- V. with Amplitude response signature attributing to assign a particular orientation*

The validation of polarization orientations follows a two-step correlation process, as illustrated in the provided figure. The piezoresponse amplitude signals (Ax, Ay, Az) are evaluated to determine the initial orientation of the regions, such as distinguishing between (100)T, (001)T, and (110)T tetragonal phases based on the presence or absence of signal components. Later, they are cross-verified with the crystallographic constraints of the $BaTiO_3$ lattice parameters and crystallographic planes observed from the XRD studies.

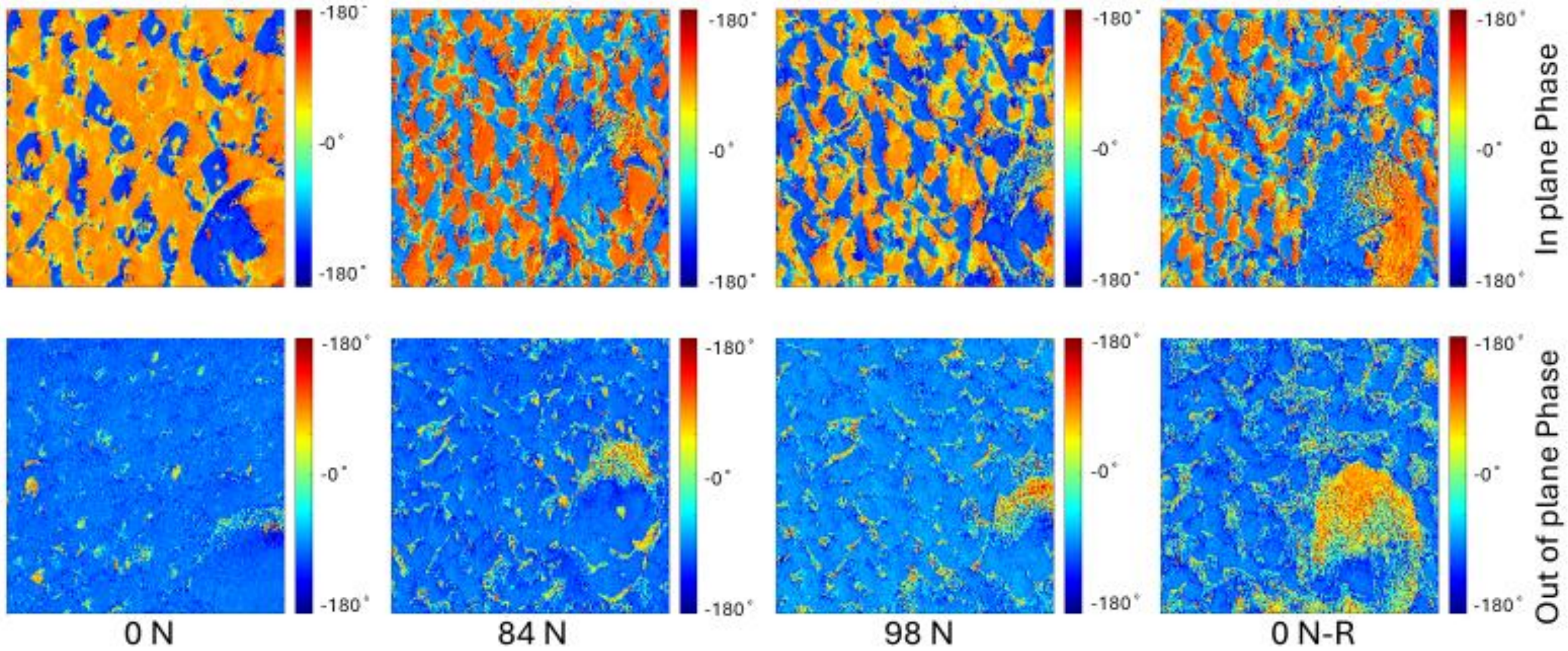

*S8. shows the in-situ PFM phase images with respect to bending loads and the evolution of domains with the application of bending loads.*

Figure S8 shows The PFM phase images in top and bottom track how the in-plane and out-of-plane contrast of ferroelectric domains change as a function of load from 0N up to 98N. The last set of images show the PFM phase image of same area on removal of the bending load. The contrast shows us the change in polarisation is irreversible beyond a certain load and is evidently seen in the retention of polarization contrast in 0N-R image.

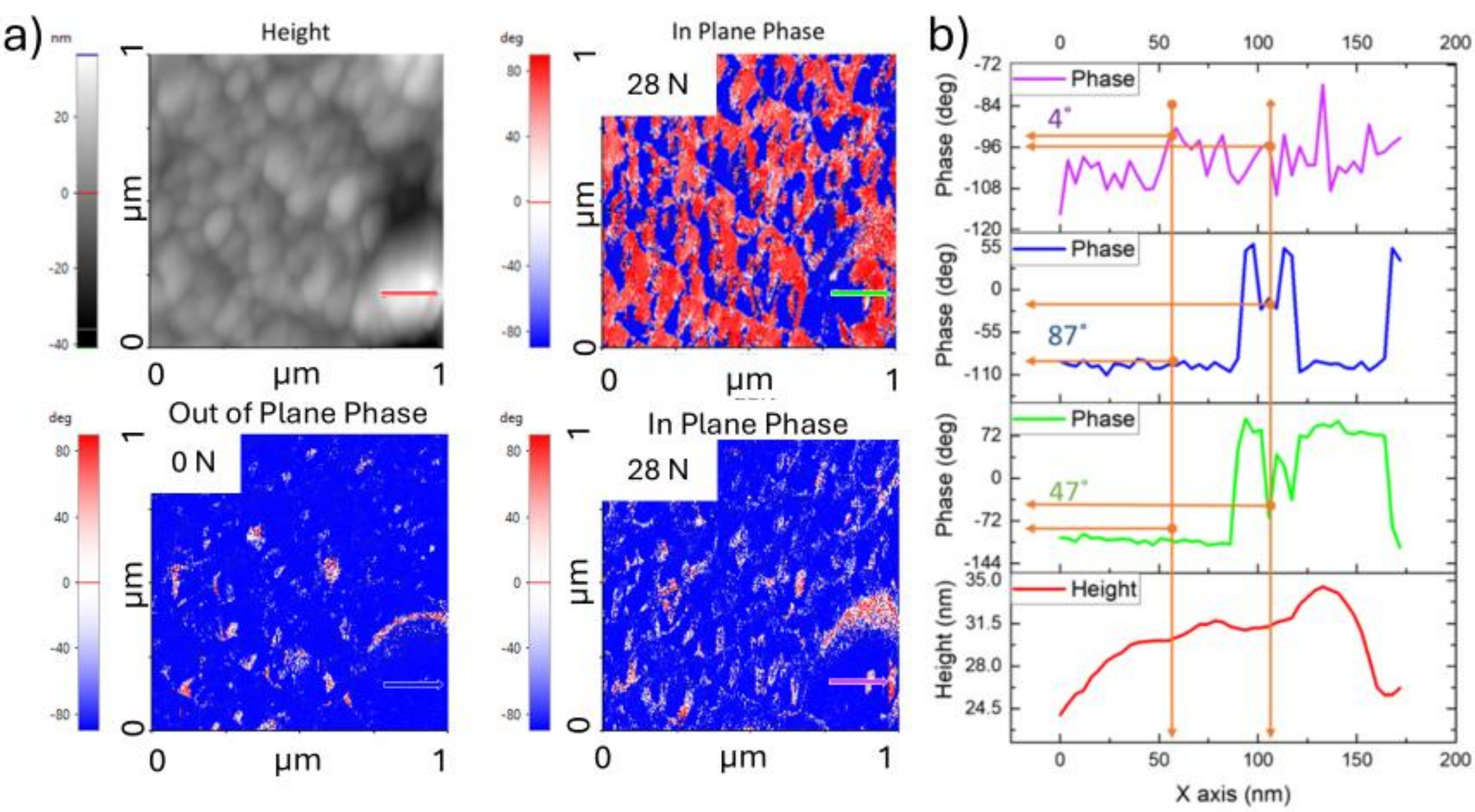

*S.9 Shows the PFM raw data of in plane and out of plane in the initial stage of bendin with 0N and 28N along with nucleation of the domains.*

Figure S9 highlights the initial nucleation of domains and the corresponding height profiles at the beginning of the bending process. And the line profile shows the nucleation is not an artifact created by path way of the cantilever motion or torsional movements.

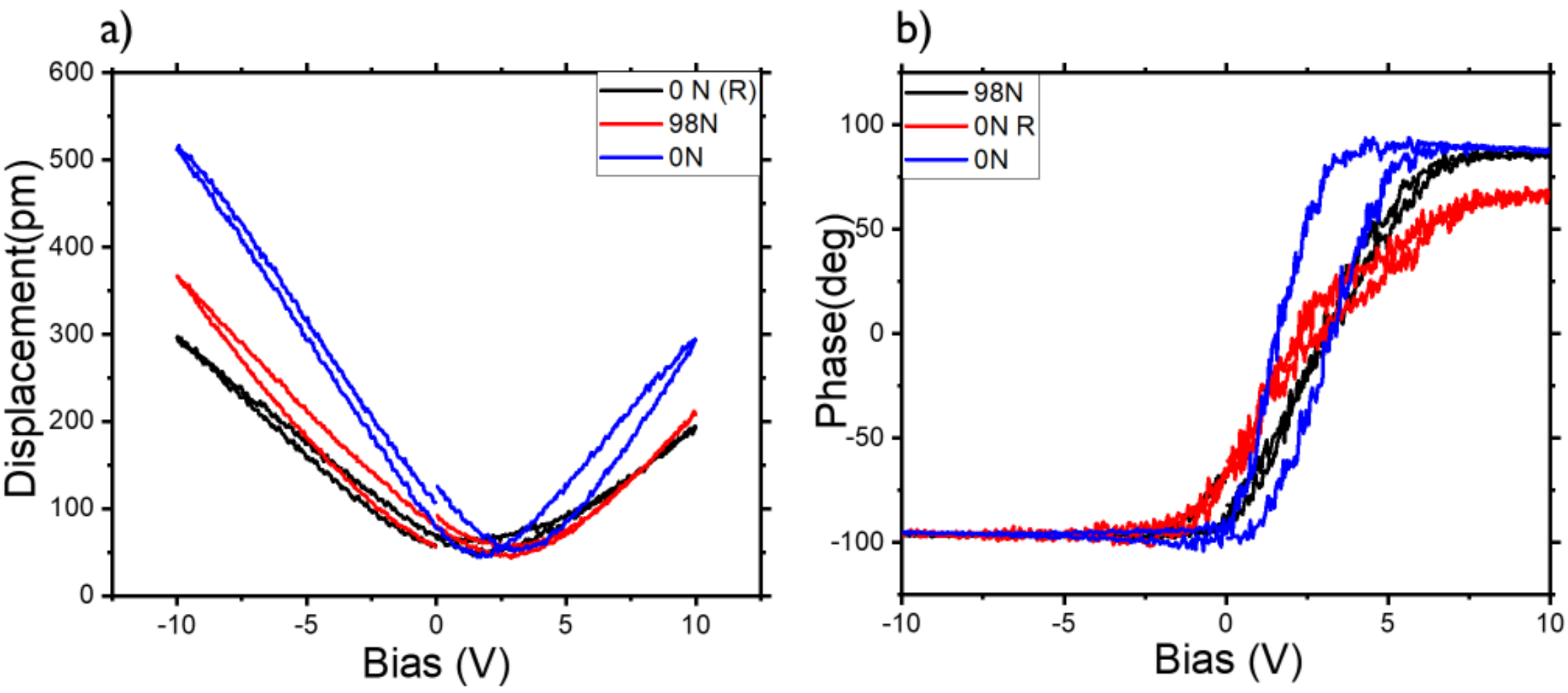

*S10. a) shows the displacement vs bias,and b) shows the phase vs bias,with respect to different bending conditions.*

Figure S10 presents PFM hysteresis loops and amplitude curves under different mechanical bending conditions. Figure S10 (a) shows the vertical displacement versus DC bias, (b) displays the typical phase vs DC bias loop, allowing for a comparison of how mechanical load affects the piezoelectric coefficient and switching barriers. A slight variation in the coercive voltages was observed under bending conditions. The pristine state indicates a relatively larger displacement plausibly due to its pre-strained growth conditions experienced by the BCZT polycrystalline thin films.

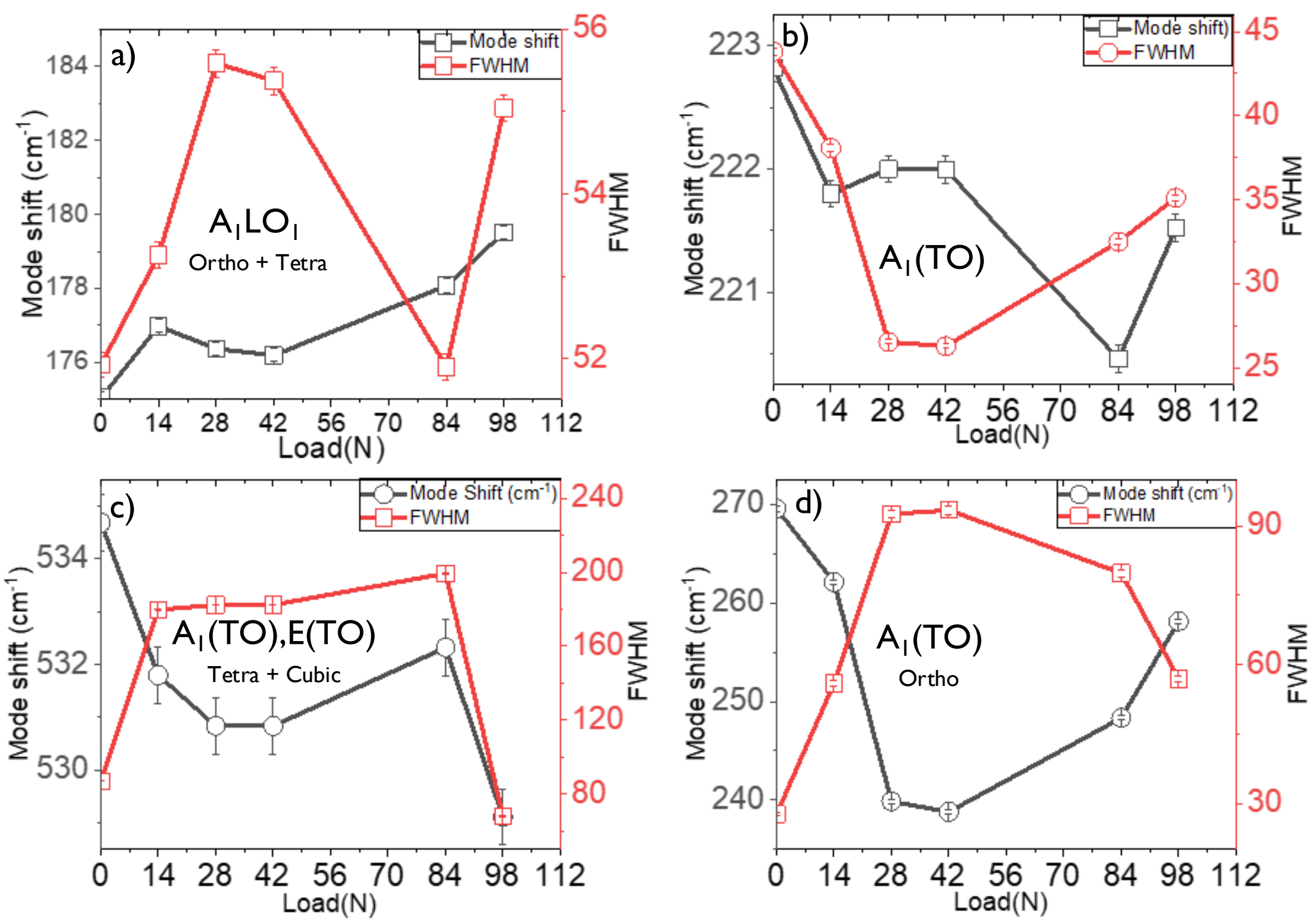

*S11. Shows the FWHM and wave number shift of different modes corresponding to BCZT phonon vibrations.*

Figure S11, shows the phonon mode changes of the BCZT thin film under varying loads studied using Raman spectroscopy. The pristine 0 N spectrum shows dominant tetragonal modes at ~530 $cm^{-1}$ and ~180 $cm^{-1}$, with weaker rhombohedral and orthorhombic contributions, indicating a tetragonal-rich state. Under 98 N bending, the rhombohedral signature at ~220 $cm^{-1}$ is suppressed; also, it could reflect $TiO_3$ vibration mode and O-Ti-O bending mode of $CaTiO_3$, which also has vibration signatures near 160 $cm^{-1}$ & 220 $cm^{-1}$; since BCZT has a higher Ca percentage, the rhombohedral ambiguity persists in these compounds. While tetragonal/orthorhombic modes sharpen and intensify, particularly in the $BO_6$ tilting (270–320 $cm^{-1}$), B–O stretching (~520 $cm^{-1}$), and $A_1g$-like (~810 $cm^{-1}$) regions, evidencing enhanced long-range order.

The plots show the Full Width at Half Maximum (FWHM) and the specific Raman mode shift for different orthorhombic and tetragonal vibration modes. a) shows Ba- O and Ca- O vibrational modes (180 $cm^{-1}$), b) and d) shows the octahedral modes of Ti-O and Zr-O (220- 250 $cm^{-1}$) and c) shows the tetragonal and cubic octahedral phonon modes of Ba-O (530 $cm^{-1}$). As the load increases from 0N to 98N, the shift observed in the phonon modes indicates the change in vibrational modes due to induced flexoelectric strain. This corroborates well with the polarization switching observed in the PFM studies. The mode shift also indicates the dominance of orthorhombic modes under bending stress conditions. In addition to emergence of orthorhombic modes, the formation of a plausible monoclinic structure could also be observed as detailed in the main text (Fig. 5).

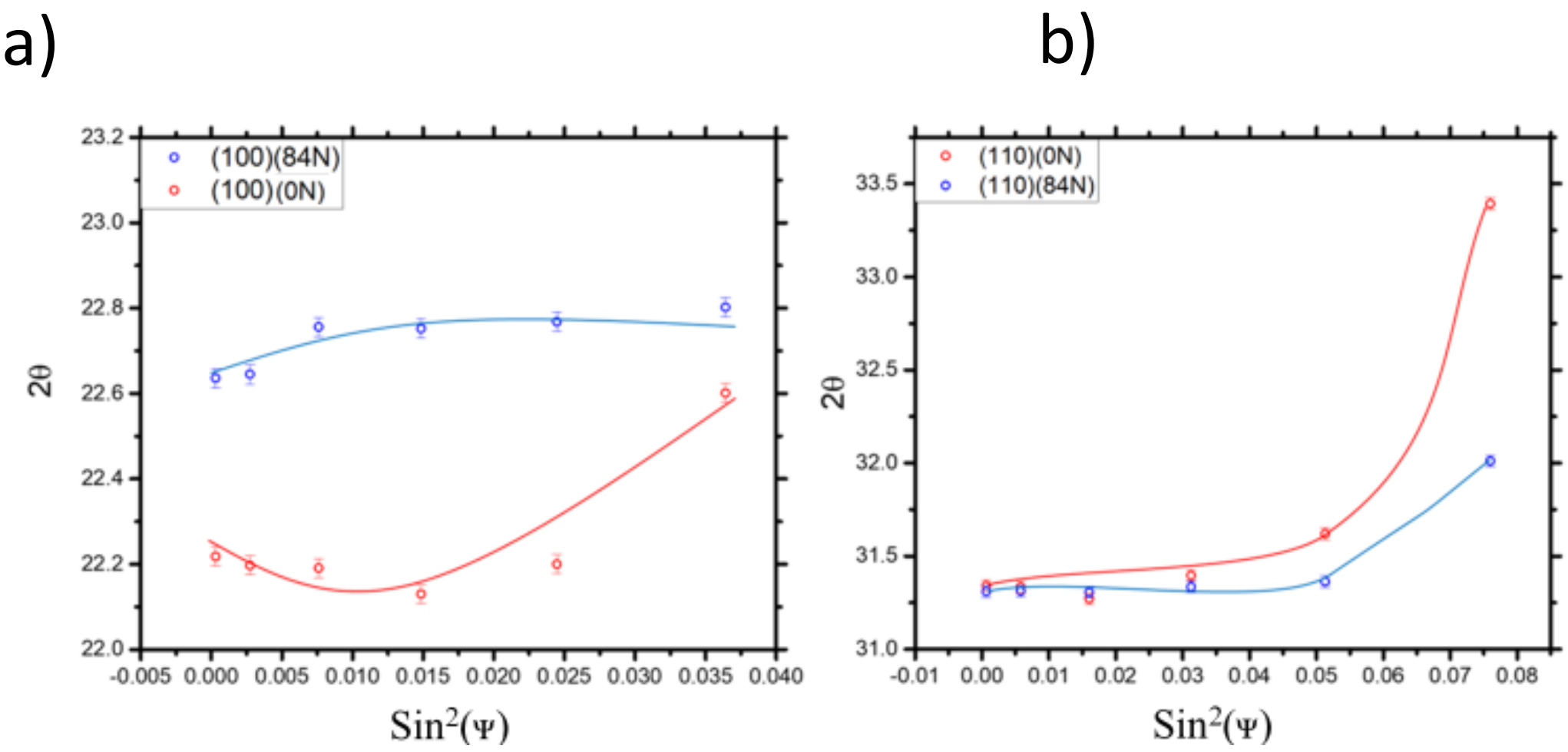

*S.12 **a** shows residual stress analysis of (100) with bending; **b** residual stress analysis of (110) with bending.*

Figure S12 shows the residual stress analysis obtained at a chosen 2θ as a function of $Sin^2(\Psi)$ ($\Psi$ - offset value of 2θ and ω). Conventionally, it is known that the slope of this curve reveals the nature of stress present in the film at a particular orientation. A positive slope indicates a compressive stress, and negative slope indicates tensile stress present in the sample. In the case of BCZT polycrystalline thin film (on (111)Pt/$TiO_2$/(001)$SiO_2$ substrate), the residual stress arising from the growth seems to be compressive in nature. However, under induced bending stress the residual stresses arerelieved from the thin film.

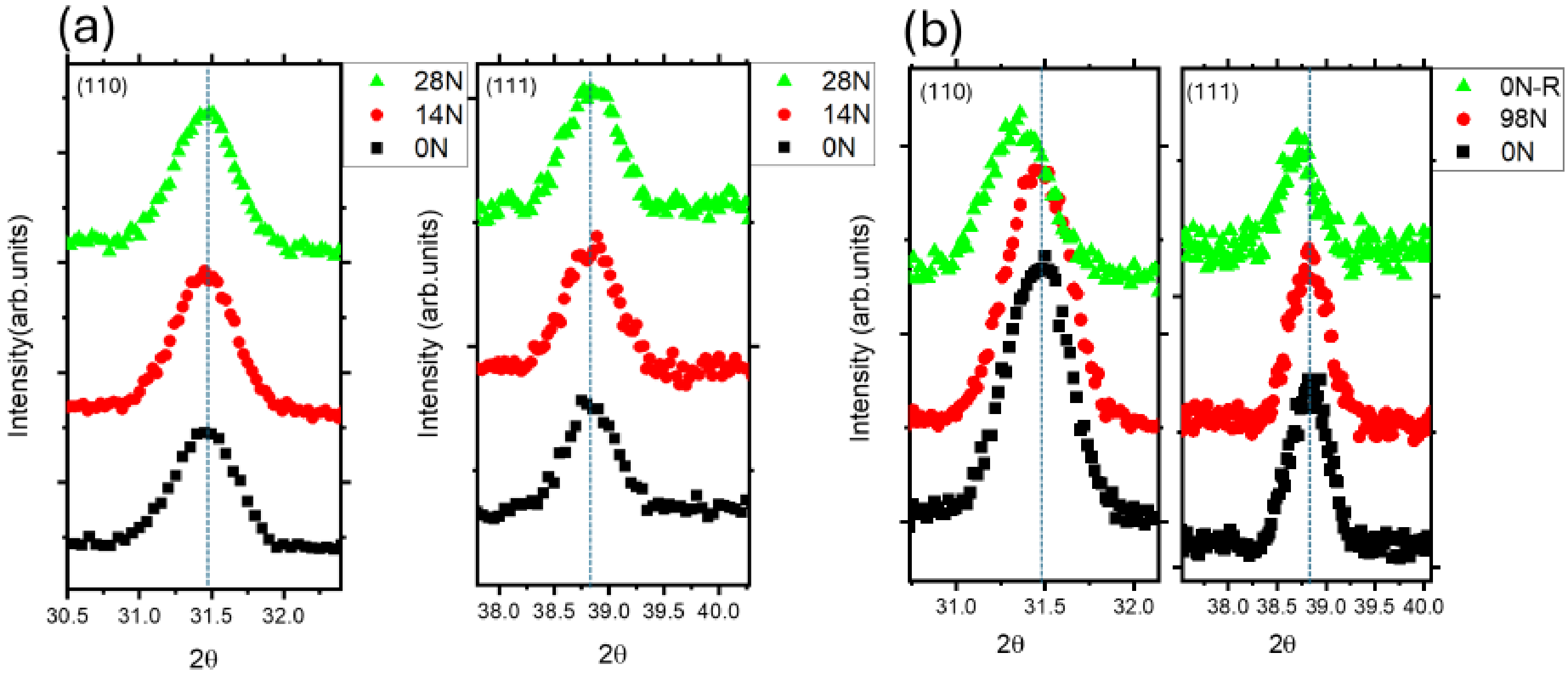

*S.13 **a** shows GIXRD with insitu bending; **b** shows GIXRD of (110) (111) peak before and after bending.*

Figures S13 shows the GIXRD of (110) and (111) diffraction peaks of BCZT polycrystalline thin film, a) showing the diffraction peak at relatively low load values b) shows the shift in peak position from the pristine state to a fully loaded state(98N) and after the complete release of the applied load(0N-R). The shift of the peak position to a lower 2theta values indicates the relaxation of the pre-existing compressive strain.